\documentclass[a4paper,11pt]{article}

\pdfoutput=1 
\usepackage{cite}
\usepackage{slashed}
\usepackage{jheppubmod}
\usepackage{enumitem}
\usepackage{amsmath}
\usepackage{tensor}
\usepackage{physics}
\usepackage{graphicx}
\usepackage{amsfonts} 
\usepackage{amssymb}
\usepackage[dvipsnames, svgnames, x11names ]{xcolor}
\usepackage{color}
\usepackage{braket}
\graphicspath{ {./1 /} }
\usepackage{subcaption}
\usepackage{graphicx}
\usepackage[T1]{fontenc} 
\usepackage{tikz}
\usepackage{physics}
\usepackage{amsmath}
\usepackage{dsfont}

\usepackage{graphicx}
\usepackage{bbold}
\graphicspath{ {./1 /} }
\colorlet{mygreen}{green!80!black}

\usepackage{tikz}
\usepackage{amsmath} 
\usepackage{mathrsfs} 
\usepackage{xfp} 
\usepackage[outline]{contour} 
\usetikzlibrary{decorations.markings,decorations.pathmorphing}
\usetikzlibrary{angles,quotes} 
\usetikzlibrary{arrows.meta} 
\contourlength{1.4pt}

\tikzset{>=latex} 
\colorlet{myred}{red!80!black}
\colorlet{myblue}{blue!80!black}
\colorlet{mygreen}{green!80!black}
\colorlet{mydarkgreen}{green!30!black}
\colorlet{mydarkred}{red!50!black}
\colorlet{myveryred}{red!30!black}
\colorlet{mydarkblue}{blue!50!black}
\colorlet{mylightblue}{mydarkblue!6}
\colorlet{mypurple}{blue!40!red!80!black}
\colorlet{mydarkpurple}{blue!40!red!50!black}
\colorlet{mylightpurple}{mydarkpurple!80!red!6}
\colorlet{myorange}{orange!40!yellow!95!black}
\tikzstyle{cone}=[mydarkblue,line width=0.2,top color=blue!60!black!30,
                  bottom color=blue!60!black!50!red!30,shading angle=60,fill opacity=0.9]
\tikzstyle{cone back}=[mydarkblue,line width=0.1,dash pattern=on 1pt off 1pt]
\tikzstyle{world line}=[myblue!60,line width=0.4]
\tikzstyle{world line t}=[mypurple!60,line width=0.4]
\tikzstyle{particle}=[mygreen,line width=0.5]
\tikzstyle{photon}=[-{Latex[length=4,width=3]},myorange,line width=0.4,decorate,
                    decoration={snake,amplitude=0.9,segment length=4,post length=3.8}]
\tikzstyle{singularity}=[myred,line width=0.6,decorate,
                         decoration={zigzag,amplitude=2,segment length=6.17}]
\tikzset{declare function={%
  penrose(\x,\c)  = {\fpeval{2/pi*atan( (sqrt((1+tan(\x)^2)^2+4*\c*\c*tan(\x)^2)-1-tan(\x)^2) /(2*\c*tan(\x)^2) )}};%
  penroseu(\x,\t) = {\fpeval{atan(\x+\t)/pi+atan(\x-\t)/pi}};%
  penrosev(\x,\t) = {\fpeval{atan(\x+\t)/pi-atan(\x-\t)/pi}};%
  kruskal(\x,\c)  = {\fpeval{asin( \c*sin(2*\x) )*2/pi}};
}}

\title{\boldmath ER=EPR and IMP wormholes in quantum gravity }

\author{Eyoab Bahiru}
\affiliation{Department of Physics, Technion,\\
Haifa 32000, Israel}

\emailAdd{joabdej@gmail.com}

\abstract{We analyze and refine the ER=EPR conjecture by studying IMP (Isenberg, Mazzeo and Pollack \cite{Isenberg:2001qe}) wormholes in QFT and quantum gravity, in the AdS/CFT context. IMP wormholes are solutions to the Einstein equations constructed from another general solution of Einstein's equations by gluing a handle on a Cauchy slice with a constant mean curvature, so that the initial data on the slice does not change in the regions away from the handle. This is made possible by using what is called the conformal method. This same method also enables one to recover a well defined perturbation theory in quantum gravity, and particularly for asymptotically AdS spacetimes \cite{Witten:2022xxp}. Thus we construct the perturbation theory on these backgrounds in addition to exploring ER=EPR when $G_{N}=0$, in the limit it is perturbatively small and beyond perturbation theory.}

\begin{document}
\maketitle
\flushbottom

\section{Introduction}
\label{sec:intro}

In physically reasonable quantum field theories, for well behaving states, for instance states with the right kind of short distance singularity, one expects that the algebra of bounded operators in a bounded (or wedge) subregions of the full spacetime are given by a type III$_{1}$ von Neumann algebra\footnote{For the proofs of such a statement in several distinct cases see \cite{bisognano1975duality,bisognano1976duality,Driessler1975,Longo1979,Fredenhagen1985,BuchholzDAntoniFredenhagen1987}}. The unique characterization of a type III$_{1}$ algebra (more precisely a factor\footnote{In general a type III$_{1}$ von Neumann algebra, on a separable Hilbert space, is a direct integral of almost all (up to a measure zero set) type III$_{1}$ factors.}) is that the entanglement spectrum across the boundary of the subregion (or the spectrum of the modular operator) is given by the set of at least all non zero real numbers, for each (normal) state. In particular, given a Cauchy slice, the entanglement across \emph{any} (sufficiently regular) co-dimension two surface that divides the slice into two is `dictated' by the type of the von Neumann algebra of operators on each side of the surface, which is type III$_{1}$ as long as the slice is smooth across the dividing surface. 

On a related note, Susskind and Maldacena conjectured \cite{MaldacenaSusskind2013} that any `subsystems' of a system described by a quantum theory of gravity like AdS/CFT that are appropriately entangled are equivalently described as being connected by an Einstein Rosen bridge, which is usually summarized as ER=EPR. The strict $G_{N}\rightarrow 0$ limit of the conjecture can be understood as concerning the entanglement between the two mouths of a slice with a wormhole. However, as we said above this entanglement across a co-dimension two surface on the wormhole neck is (mostly) described by the smoothness of the Cauchy slice across the surface and the well behaved-ness of the short distance properties of the state on the slice. Since the neck of an Einstein Rosen bridge is smooth, the entanglement across any co-dimension two surface that divides the neck is described by the type III$_{1}$ nature of the algebra of operators on each side. 

In this article one of the questions we ask is; is this the only content of the ER=EPR conjecture, when $G_{N}=0$? In particular since this statement is true for any co-dimension two surface dividing the slice. We also ask what is the content of the conjecture when perturbative and non perturbative corrections are included. The prototypical example to analyze these kinds of questions is the eternal black hole in AdS $\times C$, for compact manifold $C$ in certain kinds of theories, which is dual to a thermofield double state of two disconnected CFT's. This realizes the wormhole spacetime as a thermally entangled state of the boundary theory. For discussions on ER=EPR following the recent works on von Neumann algebras (and additional earlier works) for the large $N$ limit of the thermofield double state see \cite{EngelhardtLiu2024,BanerjeeDorbandErdmengerWeigel2023}. 

However, we proceed to analyze the issue in a different bottom up approach, of gluing a wormhole neck to the Cauchy slice with apriori no wormhole and discuss the resulting new effects in the cases when $G_{N}=0$ and when quantum gravitational interactions are included. Our starting point is the work by IMP (Isenberg, Mazzeo and Pollack \cite{Isenberg:2001qe}; and later by Isenberg, Maxwell and Pollack \cite{Isenberg:2005xp}) where they constructed Cauchy slices with wormholes in classical general relativity. The construction is modifying (performing surgery in) small subregions of a Cauchy slice which are not connected by a wormhole, so that the subregions are connected by wormholes while the metric, $g$, and the extrinsic curvature, $K$, of regions away from these sites of the surgery can be made to be close to the metric and the extrinsic curvature of the original slice arbitrarily well. This construction is done specifically on a slice with constant mean curvature ($\text{tr}_{g}K$) and uses the conformal method of constructing solutions in general relativity. Essentially the conformal method is separating $g$ and $K$ on a Cauchy slice (which is all what determines, up to differomorphism, the full solutions of Einstein's equations in a globally hyperbolic spacetime\footnote{AdS spacetime is a globally hyperbolic spacetime if one imposes the reflective boundary conditions, as is the case for AdS/CFT} as long as $g$ and $K$ satisfy the constraint equations) in to the conformal class of the metric and scale factor; and the traceless part $K$ and its trace. In the case where there is no matter and for constant mean curvature, the constraint equations become conditions on the scale factor and the restriction that the traceless part of $K$ be divergence free. Using this method IMP managed to construct, starting from a spacelike slice with data $g$ and $K$, a new topologically distinct spacelike manifold with a wormhole neck attached to the original manifold, and data $g^{'}$ and $K^{'}$ match $g$ and $K$ away from the neck. This construction works well (and in some sense better) in asymptotically AdS spacetimes, once the Cauchy slice is compactified  by including the spacelike boundary on a slice with a constant mean curvature. 

The conformal method is also essential in recovering a well defined perturbation theory in quantum gravity. We follow the relatively recent work \cite{Witten:2022xxp} on defining perturbation theory to all orders in $G_{N}$ in asymptotically AdS spacetimes, on a fixed Cauchy slice with a constant mean curvature to study these IMP wormhole spacetimes. These wormhole spacetimes could be connecting different regions of the same universe or regions of apriori disconnected slices, after a surgery on small enough balls around two generic points. We construct the QFT on these spacetimes and study the new effects arising due to the gluing of the wormhole neck that preserves the classical data far from the neck. We study these effects when perturbative corrections in $G_{N}$ are added and comment on the effects when the wormhole is Planck sized in an effort to refine the ER=EPR conjecture. 

We rephrase ER=EPR as saying ER is \emph{equivalent} to the emergence of a co-dimension two surface, $H$, that divides the neck (into two regions extending to infinity) and is stable under small gravitational perturbations, and that the algebras of operators on the neck (not the ambient region) to the left and right side of $H$ are each type III$_{1}$ when $G_{N}=0$ and type II$_{\infty}$ when $G_{N}$ is perturbatively small. These operators are dressed with respect to $H$ when we include perturbative quantum gravitational corrections, see figure \ref{fig2}. When the neck becomes Planck sized, $H$ can not be defined without an $O(1)$ ambiguity and rephrase the conjecture of ER=EPR by using the algebra of semiclassical operators in the ambient space. In particular the algebra generated by the largest algebras of operators in the ambient spaces for each of the mouths will have a non trivial commutant in $B(\mathcal{H}_{0})$ where $\mathcal{H}_{0}$ is the code subspace of the boundary Hilbert space corresponding to states with IMP wormhole. If one were to extend the operators in the ambient space to include certain heavy operators (but still with dimensions that scale slower than $N^{2}$ for the $\mathcal{N}=4$ super Yang Mills case for example), then the bridge is equivalent to these operators acting in the left and right universe and being entangled. More detailed and precise discussion will be found in section \ref{er=epr}.

In section \ref{section2}, we review the classical Einstein constraint equations, quantum constraint equations, the gluing construction of IMP wormholes and add some comments on the properties the middle region of the neck of IMP wormholes in asymptotically AdS spacetimes. In section \ref{states}, we discuss the preparation and the properties of states in the perturbative quantum gravity theory on the IMP backgrounds. In section \ref{wormcondexp}, we zoom in and focus on the regions in the middle of the neck and regions far away from the neck We discuss QFT in this background and proceed to discuss ER=EPR in the absence and presence of quantum gravitational corrections. In the last section, we discuss the full slice and discuss the perturbation theory in the full slice on both the original slice and the slice with a wormhole handle added and construct the algebras of operators. The appendices include detailed discussions on elliptic operators, the conformal method and why the degrees of freedom one gets following the conformal method are essential in recovering a well defined perturbation theory in quantum gravity.

\section{The constraint equations}\label{section2}

In standard (unitary) quantum field theories, the Hilbert space, $\mathcal{H}$, of states can be constructed by first choosing a Cauchy surface, $\Sigma$\footnote{This choice is possible for a globally hyperbolic spacetime.} and performing the path integral over all field configurations to the past of $\Sigma$ with operator insertions, and a fixed boundary conditions $\phi_{\Sigma}$ on $\Sigma$, where $\phi_{\Sigma}$'s are the dynamical coordinates\footnote{These would be the ket states and the bra vectors are constructed by a similar path integral to the future of $\Sigma$. Generally speaking taking the product of the two and integrating over $\phi_{\Sigma}$ gives the full path integral with the appropriate operator insertions; and this will be the inner product of the Hilbert space.} of the theory. 
If one hopes for something similar for gravity one will have to identify the analogous dynamical coordinates for gravity. Naively, for a $D=d+1$ dimensional spacetime, one might expect the dynamical coordinates to be the metric, $g_{ij}$, at each point on the $d$ dimensional manifold $\Sigma$. However this choice, in addition to being inconvenient for our purposes, leads to serious problems in perturbative construction of the Hilbert space. A closely related point is that classically the metric on $\Sigma$ is not arbitrary rather has to satisfy the Einstein's equations restricted to the $d$ manifold.

\subsection{The classical constraint equations}\label{gravconequ}

The Einstein constraint equations,
\begin{align}\label{fullconequ}
    D_{j}K^{j}_{k} &= D_{k}K^{j}_{j} + J_{k} \\
    R(g) &= 2\Lambda + |K|^{2}_{g} - (\text{tr}_{g} K)^{2} + 2 \rho
\end{align}
 are what are called underdetermined equations for the classical initial data $(g_{ij},K_{ij})$, where $K_{ij}$ is the extrinsic curvature. The operator $D_{i}$ and $R(g)$ are covariant derivative and the scalar curvature with respect to $g_{ij}$, $|K|^{2}_{g} = g^{ij}g^{kl}K_{ik}K_{jl}$ and $\text{tr}_{g} K = g^{ij}K_{ij}$. In addition, $J_{k} = 8\pi G\;g_{\mu k}T^{\mu}_{\nu}n^{\nu}$ and $\rho =8\pi G\; T_{\mu\nu}n^{\mu}n^{\nu}$, $n^{\mu}$ being the unit vector normal to $\Sigma$, and $T_{\mu\nu}$ being the components of the matter stress energy tensor.
 Underdetermined means that only part of the initial data is constrained, and there is another part of the data that is freely chosen. In field theories where is there is no gauge constraint, all of the data, i.e, the dynamical coordinates and their conjugate momenta are all freely chosen unlike the present case. In the case of gravity, if one is able to separate the part of the data that is fully determined by the constraints from the part that is free, then it is clear that the constraint equations will be elliptic equations (see appendix \ref{ellpequ}) for the part of the data that is fully determined. This separation of the data that is fully determined by the constraints from the part that is freely chosen can be made using what is called the conformal method, which is discussed in detail in appendix \ref{confmeth}. In the simplest case, that is in the vacuum and when the trace of the extrinsic curvature ($\tau$) is constant, the conformal method implies that the freely chosen parts of the data $(g_{ij},K_{ij})$ are the conformal class of $g_{ij}$, $\tau$, and a symmetric and traceless tensor $\sigma_{ij}$ whose divergence vanishes with respect to some arbitrary metric, which we call $\tilde{g}$, that is in the conformal class of $g_{ij}$. Then the part of the data that is fully constrained is the scale factor of the metric, $\phi$. The constraint equations will reduce to an equation called the Licnerowicz equation for $\phi$, which is elliptic as one would expect. The metric and the extrinsic curvature which satisfy the Einstein constraint equations is then given by,
\begin{align}\label{solconst}
    g_{ij}&=\phi^{-q}\tilde{g}_{ij}\\
    K_{ij}&= \phi^{-2} \sigma_{ij} + \frac{\tau}{d}\phi^{-q}\tilde{g}_{ij},
\end{align}
for $q=\frac{-4}{d-2}$. The presence of matter will only modify the term involving $\sigma_{ij}$ in the above equation, by an addition of another symmetric traceless tensor that satisfies another elliptic equation (see appendix \ref{confmeth}).

This provides the motivation to consider the conformal class of $g$ and $\tau$ to be the dynamical coordinates for gravity that we mentioned at the beginning of the section. This is indeed the starting point for \cite{York:1972sj} and the essential element in the construction of the Hilbert space of quantum gravity to all orders in perturbation theory for asymptotically AdS$_{D}$ spacetimes described in \cite{Witten:2022xxp}. In addition, this choice of dynamical coordinates will enable us to make use of a standard method of constructing initial surfaces with wormhole from any initial surface $\Sigma$ in a way that will `isolate' the effects of the wormhole handle.

\subsection{Conformal wormhole surgery}\label{confwormsurg}

One can use the freedom in the choice of the conformal class of the metric and the determined systems of equations we have been discussing so far to construct new sets of initial conditions by gluing old ones. In particular, the conformal method have been used to surgically modify a given initial data set in such a way that the new data set is the same as the original one in all but a small neighborhood of some given points \cite{joyce2002constantscalarcurvaturemetrics,Isenberg:2001qe}. The original initial data sets can be on a connected or disconnected manifolds. As alluded to before, for us the kind of surgery we are interested in is the one where a handle is added to a given manifold without modifying initial data in regions away from the handle. We may also have two disconnected components of the manifold and a handle (or a neck) is added to produce an initial data on the connected sum of the two components, see figure \ref{fig2}.  

\providecolor{mydarkblue}{RGB}{20,80,140}   
\providecolor{myred}{RGB}{190,30,45}
\colorlet{slcA}{mydarkblue!28}
\colorlet{slcB}{myred!28}

\newcommand{\wormneck}[9]{%
  \begin{scope}
    \pgfmathsetmacro{\xm}{(#1+#2)/2}
    \pgfmathsetmacro{\yo}{#3+#6}          
    \pgfmathsetmacro{\yi}{#3+#6-2.1*#4}   
    \pgfmathsetmacro{\kk}{0.62*(#2-#1)/2} 
    \path[left color=#7, right color=#8, draw=black, line width=0.7pt, line join=round]
      (#1-#4,#3) .. controls +(0,0.55*#6) and +(-\kk-#4,0) .. (\xm,\yo)
                 .. controls +(\kk+#4,0) and +(0,0.55*#6) .. (#2+#4,#3)
      arc[start angle=0,end angle=-180,x radius=#4,y radius=#5]
                 .. controls +(0,0.5*#6) and +(\kk-#4,0) .. (\xm,\yi)
                 .. controls +(-\kk+#4,0) and +(0,0.5*#6) .. (#1+#4,#3)
      arc[start angle=0,end angle=-180,x radius=#4,y radius=#5] -- cycle;
    \pgfmathsetmacro{\ry}{(\yo-\yi)/2}
    \draw[densely dashed, line width=0.4pt, black!55]
      (\xm,\yo) arc[start angle=90,end angle=-90,x radius=0.32*\ry,y radius=\ry];
    \draw[line width=0.6pt]
      (\xm,\yo) arc[start angle=90,end angle=270,x radius=0.32*\ry,y radius=\ry];
    \node[font=\small, above=1pt] at (\xm,\yo) {#9};
  \end{scope}}

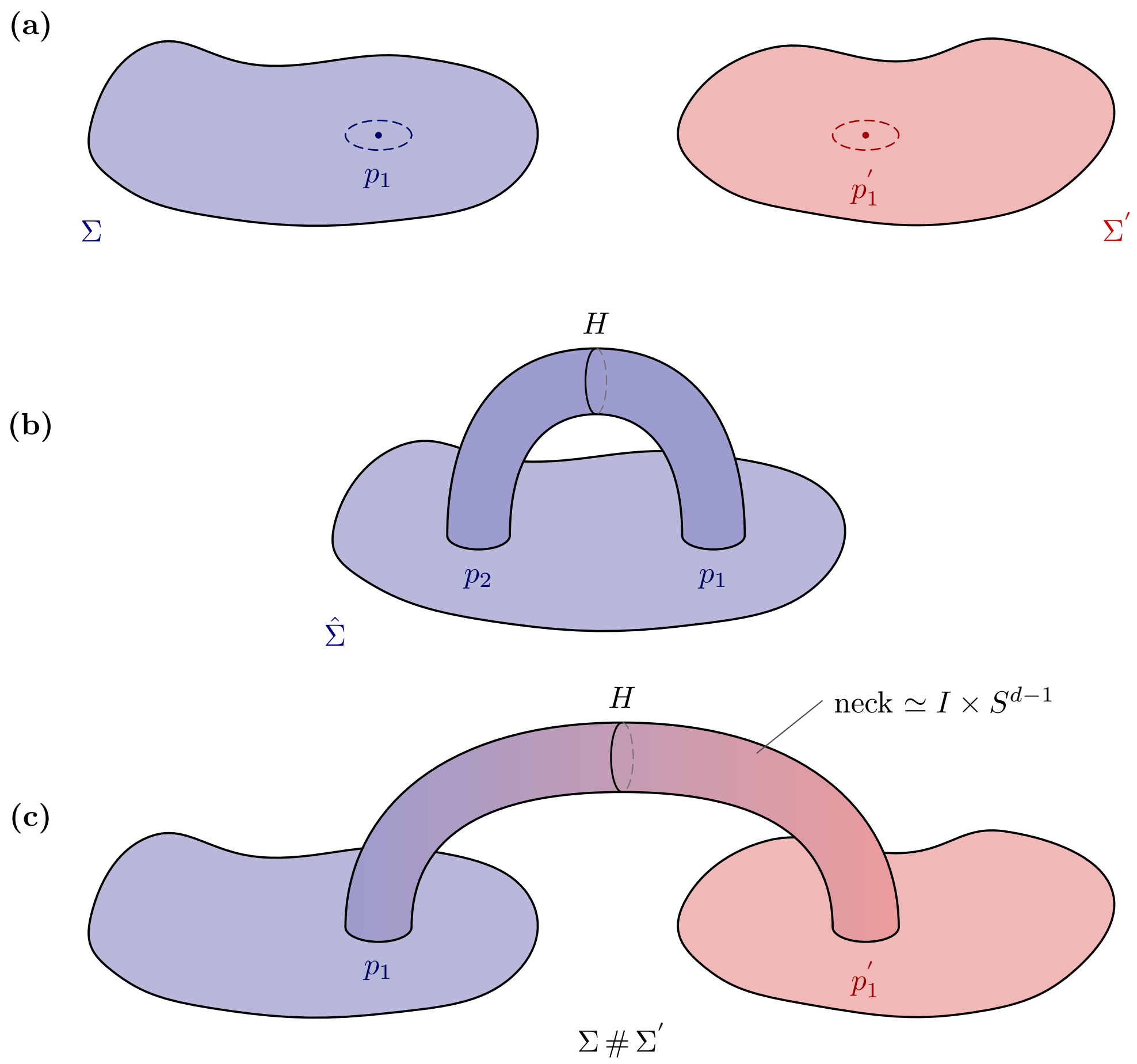
\begin{figure}[t]
\centering
\begin{tikzpicture}[x=1cm,y=1cm,line cap=round,line join=round,
    slice/.style={draw=black, line width=0.7pt},
    lbl/.style={font=\small},
    panel/.style={font=\small\bfseries}]

\begin{scope}
  \node[panel] at (-6.6,1.2) {(a)};
  \path[slice, fill=slcA]
    plot[smooth cycle, tension=0.75] coordinates
    {(-5.9,0.1) (-5.2,1.0) (-3.9,0.75) (-2.2,0.85) (-0.9,0.35)
     (-1.1,-0.65) (-2.6,-1.05) (-4.4,-1.0) (-5.6,-0.6)};
  \path[slice, fill=slcB]
    plot[smooth cycle, tension=0.75] coordinates
    {(0.9,0.2) (1.9,0.95) (3.4,0.8) (4.6,1.05) (5.8,0.45)
     (5.4,-0.55) (4.1,-1.05) (2.4,-0.95) (1.2,-0.55)};
  \foreach \x/\c/\n in {-2.6/mydarkblue/p_{1}, 3.0/myred/p_{1}^{'}}{
    \draw[densely dashed, line width=0.5pt, \c!80!black]
      (\x,-0.05) ellipse[x radius=0.38,y radius=0.17];
    \fill[\c!80!black] (\x,-0.05) circle[radius=1.1pt];
    \node[lbl, \c!80!black, below=3pt] at (\x,-0.2) {$\n$};
  }
  \node[lbl, mydarkblue] at (-5.9,-1.15) {$\Sigma$};
  \node[lbl, myred]      at (5.9,-1.1)  {$\Sigma^{'}$};
\end{scope}

\begin{scope}[yshift=-4.6cm]
  \node[panel] at (-6.6,1.2) {(b)};
  \path[slice, fill=slcA]
    plot[smooth cycle, tension=0.75] coordinates
    {(-3.1,0.1) (-2.3,1.0) (-0.9,0.8) (1.0,0.9) (2.6,0.4)
     (2.4,-0.65) (0.8,-1.1) (-1.2,-1.05) (-2.7,-0.6)};
  \wormneck{-1.45}{1.25}{-0.05}{0.36}{0.16}{2.15}
           {slcA!85!mydarkblue}{slcA!85!mydarkblue}{$H$}
  \node[lbl, mydarkblue] at (-3.1,-1.15) {$\hat{\Sigma}$};
  \node[lbl, mydarkblue!80!black, below=3pt] at (-1.45,-0.2) {$p_{2}$};
  \node[lbl, mydarkblue!80!black, below=3pt] at (1.25,-0.2) {$p_{1}$};
\end{scope}

\begin{scope}[yshift=-9.1cm]
  \node[panel] at (-6.6,1.2) {(c)};
  \path[slice, fill=slcA]
    plot[smooth cycle, tension=0.75] coordinates
    {(-5.9,0.1) (-5.2,1.0) (-3.9,0.75) (-2.2,0.85) (-0.9,0.35)
     (-1.1,-0.65) (-2.6,-1.05) (-4.4,-1.0) (-5.6,-0.6)};
  \path[slice, fill=slcB]
    plot[smooth cycle, tension=0.75] coordinates
    {(0.9,0.2) (1.9,0.95) (3.4,0.8) (4.6,1.05) (5.8,0.45)
     (5.4,-0.55) (4.1,-1.05) (2.4,-0.95) (1.2,-0.55)};
  \wormneck{-2.6}{3.0}{-0.05}{0.38}{0.17}{2.35}
           {slcA!85!mydarkblue}{slcB!85!myred}{$H$}
  \node[lbl, mydarkblue!80!black, below=3pt] at (-2.6,-0.2) {$p_{1}$};
  \node[lbl, myred!80!black, below=3pt] at (3.0,-0.2) {$p_{1}^{'}$};
  \node[lbl] at (0.2,-1.35) {$\Sigma\,\#\,\Sigma^{'}$};
  \draw[line width=0.4pt, black!70] (1.75,1.95) -- (2.5,2.55)
    node[right, lbl, text=black] {neck $\simeq I\times S^{d-1}$};
\end{scope}
\end{tikzpicture}
\caption{(a) Two disconnected Cauchy slices $\Sigma$ and $\Sigma^{'}$, with the points $p_{1}$, $p_{1}^{'}$
around which small balls (dashed, with boundary $S^{d-1}$) are excised.
(b) A handle added to a single slice, giving $\hat{\Sigma}$: two balls are excised from
$\Sigma$ and their boundaries are joined by a neck $I\times S^{d-1}$, where $I$ is an interval.
(c) The connected sum $\Sigma\,\#\,\Sigma^{'}$ connects the two disconnected slices: the two slices are joined by a neck;
$H$ is the minimal sphere at the middle of the neck, which will be relevant later.}
\label{fig2}
\end{figure}

We start with a simpler discussion of the vacuum case, then later discuss the changes that will arise in the presence of a matter field. If we start with some spacelike manifold $\Sigma$ (or $\Sigma \cup \Sigma^{'}$) with the initial condition $(g,K)$ that satisfy the constraint equations with negative cosmological constant, we assume that $\tau = \text{tr}_{g}K=0$ however it is important that $K \neq 0 $ for this technique to work in flat space (when there is a non zero cosmological constant, this is possible \cite{Isenberg:2005xp}). Then we consider another spacelike manifold $\hat{\Sigma}$ (or $\Sigma \# \Sigma^{'}$) where a  handle or a neck is added. Using the above conformal method, we can construct initial data sets on this manifold too, however, the non trivial question is whether or not we can find an initial condition that approximates the original initial condition in the region away from the wormhole arbitrarily well. In the following sections, the original and the new initial conditions will correspond to the two backgrounds on which we will define the perturbative Hilbert spaces.

The way we do this is as follows; starting with $\Sigma$ (or $\Sigma \cup \Sigma^{'}$), we pick any two generic points\footnote{For technical reasons there should not be any non trivial conformal Killing vector fields that vanish at these points; and this condition is satisfied for a generic choice of the points \cite{Isenberg:2001qe}.} (one on each component in the disconnected case) around which we consider a small neighborhood of the points at most a geodesic distance $R$ away. We then remove the two points from the manifold and conformally map the two punctured balls to an asymptotically cylindrical neighborhoods where the removed points correspond to the asymptotic infinity of the cylinders. This is possible because of a simple property of Riemannian manifolds that in balls such as these, the radial geodesic distance directions meet with the spherical, $S^{d-1}$, directions perpendicularly, thus we can write the metric in each of these balls\footnote{This property goes by the name Gauss Lemma in geometric analysis.} as,
\begin{equation}
    ds_{\Sigma}^{2}|_{B} = dr^{2} + r^{2}h_{ab}(r)d\theta^{a}d\theta^{b}
\end{equation}
 where $r$ is the geodesic distance inside the balls from each of the corresponding points, while $h_{ab}(r)$ and $\theta^{a}$ are the metric and coordinates on the sphere $S^{d-1}$ at distance $r$. Note that, even though for notational ease we have only written one line element above, there are two such metrics and the coordinates which are distinct, corresponding to each of the balls around the generically chosen points. If the metric $g$ were conformally flat, $h_{ab}(r)$ would be just the coefficients of the standard metric on sphere, $d\Omega_{d-1}$. However, in general, all we can say is that it should go smoothly to $d\Omega_{d-1}$ as $r \rightarrow 0$. In particular, we have $h_{ab}(r)d\theta^{a}d\theta^{b} = d\Omega_{d-1} + r\; k_{ab}(r)d\theta^{a}d\theta^{b}$, for some metric $k_{ab}$ that is smooth in $r$. 

We conformally map the metric $g|_{B}$ inside the balls to get a new metric, $\tilde{g}|_{B}$, by multiplication with $r^{-2}$. Extending this to the full slice, we can consider a Weyl transformation of the full metric like \eqref{solconst} with,
\begin{equation}\label{balltocylinder}
   \phi = \begin{cases}
          &r^{-2/q}, \;\text{ inside } B_{1,2} \\
          &f(r) , \;\text{ inside } B^{0}_{1,2}\text{\textbackslash} B_{1,2} \\
          & 1, \;\text{ elsewhere }
          \end{cases}
\end{equation}
where $B^{0}$ a ball of radius $R_{0}=R+\Delta R$ and $f$ is a smooth function that interpolates between $B$ and the part of the Cauchy slice outside $B^{0}$, with $f(R)=R^{-2/q}$ and $f(R_{0}) = 1$. Then the new metric $\tilde{g}$ defined on the full slice by a Weyl transformation of $g$ with $\phi^{q}$ given above, will be in the conformal class of $g$. In addition, defining a two tensor $\sigma$ by multiplying the extrinsic curvature of the initial data with $\phi^{2}$, we get a new set of initial conditions $(\tilde{g},\sigma)$ which do not themselves satisfy the constraint equations but are related to the ones that do with Weyl transformation with $\phi$ just like in \eqref{solconst}.  

It is easy to see that this new metric $\tilde{g}$ is asymptotically cylindrical if we define a new coordinate $z=-\; \text{log } r + \log R$ in the range $0\leq z < \infty$, then we have;
\begin{equation}
     d\tilde{s}_{\Sigma}^{2}|_{B} = dz^{2} + d\Omega_{d-1} + e^{-z}\; k_{ab}(r)d\theta^{a}d\theta^{b}
\end{equation}
where the points we started with (which are at $r=0$ by definition) are mapped to $z=\infty$.

The goal is to identify some large but finite, $0 \leq z \leq T$ , length of the two cylinders (which we will call $N_{i,T}$ where $i=1,2$). But to have the correct identification we first map $z \rightarrow -z + T$ and points on the sphere to their reflection counter-points with respect to the origin (as one would expect from the orientation change following as we `construct' the handle), for one of the two cylinders. Then we identify the coordinates of the two cylinders (see figure \ref{fig3}) to create the wormhole handle. This will give us a new manifold $\Sigma_{T}$ topologically isomorphic to $\hat{\Sigma}$ (or $\Sigma\#\Sigma^{'}$ if the original manifolds were disconnected).

The metric $\tilde{g}_{T}$ on this new manifold will be the same as the original one in the regions outside the balls. In the wormhole region, it will be constructed out of the metrics in each of the two cylindrical regions, $\tilde{g}_{T}|_{N_{1,T}}$ and $\tilde{g}_{T}|_{N_{2,T}}$. We have said that the handle is topologically $I\cross S^{d-1}$, and we can divide the handle into three parts, two parts $E_{1}$ and $E_{2}$, close to each of the mouths of wormhole and an intermediate part we call $E_{0}$. Now we consider two smooth functions $\chi_{1,2}$ defined on the handle but their sum always equals one at any point on the handle. We take them to have the following property; $\chi_{1}$ is one in $E_{1}$ and zero in $E_{2}$, while $\chi_{2}$ is one in $E_{2}$ and zero in $E_{1}$. In the intermediate region $E_{0}$, both functions are non zero while smoothly connecting their values in $E_{1}$ and $E_{2}$, while preserving the condition that they sum to one. Then the metric on the handle is given by
\begin{equation}\label{partglu}
    \tilde{g}_{T}|_{N_{T}} = \chi_{1} \tilde{g}|_{N_{1,T}} + \chi_{2} \tilde{g}|_{N_{2,T}}
\end{equation}
where $\tilde{g}_{T}|_{N_{i,T}}$, with $i=1,2$, are the metric restricted to each of the cylinders. Similarly we can construct a symmetric traceless tensor\footnote{This is possible with the appropriate choice of the partitions, $\chi_{i}$'s.} $\sigma_{T}$ on the manifold, which is the same as $\sigma$ outside the balls and on the wormhole region is given by $\sigma_{T}|_{N} = \chi_{1} \sigma|_{N_{1,T}} + \chi_{2} \sigma|_{N_{2,T}}$. Again $\sigma|_{N_{i,T}}$ are the TT two tensors restricted to each of the cylinders.

This new initial data $(\tilde{g}_{T},\sigma_{T})$ on $\Sigma_{T}$ will in general not satisfy the constraint equations. However, once we make $\sigma_{T}$ transverse with respect to $\tilde{g}_{T}$, we can hope to find a $\phi_{T}$, which will determine for us the correct initial data that satisfy the constraint equations in the Weyl orbit of $\tilde{g}_{T}$ and the modified $\sigma_{T}$. The central result of \cite{Isenberg:2001qe} is that this is indeed possible with a modification of $\sigma_{T}$ that gives an initial data that is arbitrarily close to the original data. That is, once they found the modified TT tensor, $\sigma_{T} - \mu_{T}$ where $\mu_{T}$ is small in a precise sense, they were able to find the appropriate $\phi_{T}$, constructed by adding a small correction to the original $\phi$ discussed above and its gluing in the wormhole region, where 
\begin{align}
    g_{T} &= \phi_{T}^{-q}\tilde{g}_{T}\\
    K_{T} &= \phi_{T}^{-2} (\sigma_{T} - \mu_{T})
\end{align}
satisfy the Einstein constraint equations. In addition, they showed rigorously that the initial conditions $(g_{T},K_{T})$ approximates the original initial conditions $(g,K)$ arbitrary well as $T\rightarrow \infty$ outside the wormhole region.  This follows from precise estimates of the norms of the modification in $\phi_{T}$, and $\mu_{T}$ in the appropriate spaces. The exact computation was done in \cite{Isenberg:2001qe} for the case of zero cosmological constant but the same goes for negative cosmological constant. This is because the only place where the cosmological constant enters the constraints is in the Hamiltonian constraint, i.e, the Licnerowicz equation, whose only relevant aspect used in the computation is the positivity of $\beta$ (see appendix \ref{confmeth}), which will still be true when a negative cosmological constant is present. This is also the reason why the conformal method is well suited for spacetimes with negative or zero cosmological constants, for which the Licnerowicz equation is better behaved.

\providecolor{mydarkblue}{RGB}{20,80,140}  
\providecolor{myred}{RGB}{190,30,45}

\newcommand{\gcAB}{0.90}  
\newcommand{\gcBB}{0.36}  
\newcommand{\gcAZ}{1.45}  
\newcommand{\gcBZ}{0.58}  
\newcommand{\gcA}{0.50}   
\newcommand{\gcB}{0.20}   
\newcommand{\gcYq}{1.50}  
\newcommand{\gcYh}{2.20}  
\newcommand{\gcYT}{4.00}  
\newcommand{\gcYtop}{5.20}

\newcommand{\gluesheet}{%
  \path[draw=black, line width=0.7pt, fill=mydarkblue!14, line join=round]
    (-6.6,-1.25) .. controls (-3.0,-1.55) and (2.5,-0.95) .. (5.9,-1.25)
    -- (6.6,0.85) .. controls (3.0,1.10) and (-2.5,0.60) .. (-5.9,0.85) -- cycle;
  \node[font=\small, mydarkblue] at (-5.85,-0.95) {$\Sigma$};}

\newcommand{\gluecyl}[2]{%
  \begin{scope}[shift={(#1,#2)}]
    \draw[densely dashed, line width=0.4pt, black!50]
      (-\gcAB,0) arc[start angle=180,end angle=0,x radius=\gcAB,y radius=\gcBB];
    \path[left color=black!22, right color=black!22, middle color=black!4]
      (-\gcA,\gcYT) -- (-\gcA,\gcYtop)
      arc[start angle=180,end angle=0,x radius=\gcA,y radius=\gcB]
      -- (\gcA,\gcYT) arc[start angle=0,end angle=-180,x radius=\gcA,y radius=\gcB] -- cycle;
    \draw[line width=0.6pt, black!55] (-\gcA,\gcYT) -- (-\gcA,\gcYtop) (\gcA,\gcYT) -- (\gcA,\gcYtop);
    \draw[line width=0.6pt, black!55, dash pattern=on 1pt off 2.2pt]
      (-\gcA,\gcYtop) -- ++(0,0.50) (\gcA,\gcYtop) -- ++(0,0.50);
    \path[draw=black!55, line width=0.6pt, fill=black!12]
      (0,\gcYtop) ellipse[x radius=\gcA,y radius=\gcB];
    \path[left color=mydarkblue!42, right color=mydarkblue!42, middle color=mydarkblue!9]
      (-\gcAB,0) .. controls (-\gcA-0.03,0.22) and (-\gcA,0.45) .. (-\gcA,0.95)
      -- (-\gcA,\gcYT) arc[start angle=180,end angle=360,x radius=\gcA,y radius=\gcB]
      -- (\gcA,0.95) .. controls (\gcA,0.45) and (\gcA+0.03,0.22) .. (\gcAB,0)
      arc[start angle=0,end angle=-180,x radius=\gcAB,y radius=\gcBB] -- cycle;
    \draw[line width=0.7pt, line cap=round]
      (-\gcAB,0) .. controls (-\gcA-0.03,0.22) and (-\gcA,0.45) .. (-\gcA,0.95) -- (-\gcA,\gcYT)
      (\gcAB,0)  .. controls (\gcA+0.03,0.22)  and (\gcA,0.45)  .. (\gcA,0.95)  -- (\gcA,\gcYT);
    \draw[line width=0.7pt]
      (-\gcAB,0) arc[start angle=180,end angle=360,x radius=\gcAB,y radius=\gcBB];
    \foreach \h in {\gcYq,\gcYT}{
      \draw[densely dashed, line width=0.4pt, black!50]
        (-\gcA,\h) arc[start angle=180,end angle=0,x radius=\gcA,y radius=\gcB];
      \draw[line width=0.6pt]
        (-\gcA,\h) arc[start angle=180,end angle=360,x radius=\gcA,y radius=\gcB];}
    \draw[densely dashed, line width=0.5pt, myred!70]
      (-\gcA,\gcYh) arc[start angle=180,end angle=0,x radius=\gcA,y radius=\gcB];
    \draw[line width=1.1pt, myred]
      (-\gcA,\gcYh) arc[start angle=180,end angle=360,x radius=\gcA,y radius=\gcB];
  \end{scope}}

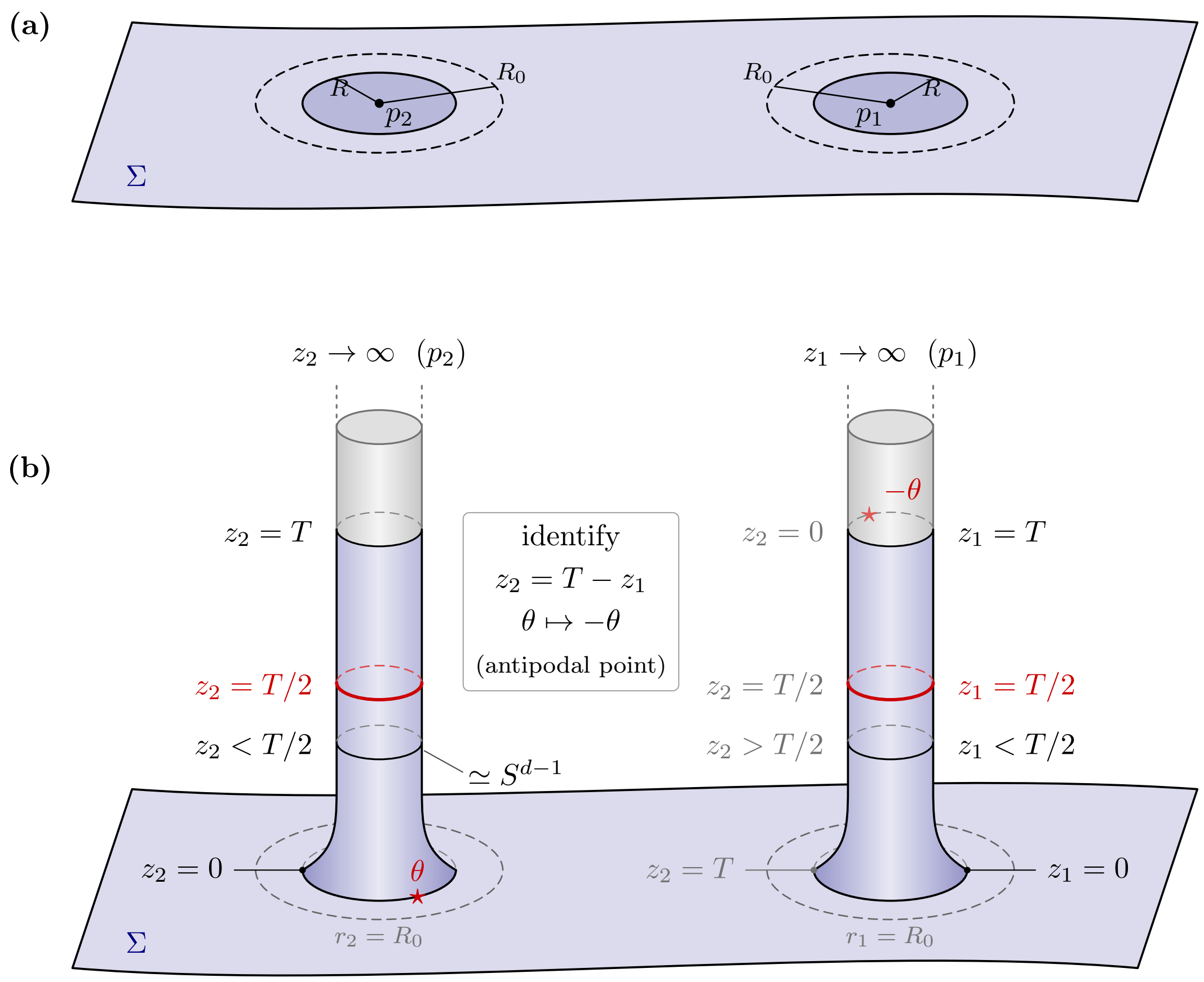
\begin{figure}[t]
\centering
\begin{tikzpicture}[x=1cm,y=1cm,line cap=round,line join=round,
    lbl/.style={font=\small}, glbl/.style={font=\small, text=black!55},
    panel/.style={font=\small\bfseries}]

\begin{scope}
  \node[panel] at (-7.1,0.8) {(a)};
  \gluesheet
  \foreach \x/\n/\aR/\aRz/\s in {-3.0/2/125/20/-1, 3.0/1/55/160/1}{
    \begin{scope}[shift={(\x,-0.1)}]
      \path[fill=mydarkblue!28, draw=black, line width=0.7pt]
        (0,0) ellipse[x radius=\gcAB,y radius=\gcBB];
      \draw[densely dashed, line width=0.6pt] (0,0) ellipse[x radius=\gcAZ,y radius=\gcBZ];
      \draw[line width=0.5pt] (0,0) -- ({\gcAB*cos(\aR)},{\gcBB*sin(\aR)});
      \draw[line width=0.5pt] (0,0) -- ({\gcAZ*cos(\aRz)},{\gcBZ*sin(\aRz)});
      \fill (0,0) circle[radius=1.5pt];
      \node[lbl, inner sep=1pt] at ({-0.24*\s},-0.17) {$p_{\n}$};
      \node[font=\scriptsize, inner sep=1pt]
        at ({0.55*\gcAB*cos(\aR)+0.19*\s},{0.55*\gcBB*sin(\aR)+0.02}) {$R$};
      \node[font=\scriptsize, inner sep=1pt]
        at ({1.14*\gcAZ*cos(\aRz)},{1.14*\gcBZ*sin(\aRz)+0.13}) {$R_{0}$};
    \end{scope}}
\end{scope}
\begin{scope}[yshift=-9cm]
  \node[panel] at (-7.1,4.6) {(b)};
  \gluesheet
  \foreach \x in {-3.0,3.0}
    \draw[densely dashed, line width=0.5pt, black!60]
      (\x,-0.1) ellipse[x radius=\gcAZ,y radius=\gcBZ];
  \gluecyl{-3.0}{-0.1}
  \gluecyl{3.0}{-0.1}

  \begin{scope}[shift={(-3.0,-0.1)}]
    \node[lbl, anchor=east] (zl) at (-\gcAZ-0.25,0.0) {$z_{2}=0$};
    \draw[line width=0.4pt] (zl.east) -- (-\gcAB,0);       
    \fill (-\gcAB,0) circle[radius=1.1pt];
    \node[font=\scriptsize, text=black!55] at (0,-\gcBZ-0.20) {$r_{2}=R_{0}$};
    \node[lbl, anchor=east] at (-\gcA-0.15,\gcYq-0.05)       {$z_{2}<T/2$};
    \node[lbl, anchor=east, myred] at (-\gcA-0.15,\gcYh-0.05) {$z_{2}=T/2$};
    \node[lbl, anchor=east] at (-\gcA-0.15,\gcYT-0.05)       {$z_{2}=T$};
    \node[lbl, anchor=south] at (0,\gcYtop+0.55)             {$z_{2}\to\infty\ \ (p_{2})$};
    \draw[line width=0.4pt, black!70] (\gcA+0.03,\gcYq-0.10) -- ++(0.45,-0.25)
      node[lbl, right, text=black, inner sep=1.5pt] {$\simeq S^{d-1}$};
    \node[myred, inner sep=0pt] (sL)
      at ({\gcAB*cos(-60)},{\gcBB*sin(-60)}) {\small$\boldsymbol{\star}$};
    \node[lbl, myred, anchor=south, inner sep=2.5pt] at (sL.north) {$\theta$};
  \end{scope}

  \begin{scope}[shift={(3.0,-0.1)}]
    \node[lbl, anchor=west] (zr) at (\gcAZ+0.25,0.0) {$z_{1}=0$};
    \draw[line width=0.4pt] (zr.west) -- (\gcAB,0);         
    \fill (\gcAB,0) circle[radius=1.1pt];
    \node[font=\scriptsize, text=black!55] at (0,-\gcBZ-0.20) {$r_{1}=R_{0}$};
    \node[lbl, anchor=west] at (\gcA+0.15,\gcYq-0.05)        {$z_{1}<T/2$};
    \node[lbl, anchor=west, myred] at (\gcA+0.15,\gcYh-0.05)  {$z_{1}=T/2$};
    \node[lbl, anchor=west] at (\gcA+0.15,\gcYT-0.05)        {$z_{1}=T$};
    \node[lbl, anchor=south] at (0,\gcYtop+0.55)             {$z_{1}\to\infty\ \ (p_{1})$};
    \node[glbl, anchor=east] (zrl) at (-\gcAZ-0.25,0.0) {$z_{2}=T$};
    \draw[line width=0.4pt, black!55] (zrl.east) -- (-\gcAB,0);
    \fill[black!55] (-\gcAB,0) circle[radius=1.1pt];
    \node[glbl, anchor=east] at (-\gcA-0.15,\gcYq-0.05)      {$z_{2}>T/2$};
    \node[glbl, anchor=east] at (-\gcA-0.15,\gcYh-0.05)      {$z_{2}=T/2$};
    \node[glbl, anchor=east] at (-\gcA-0.15,\gcYT-0.05)      {$z_{2}=0$};
    \node[myred!65, inner sep=0pt] (sR)
      at ({\gcA*cos(120)},{\gcYT+\gcB*sin(120)}) {\small$\boldsymbol{\star}$};
    \node[lbl, myred, anchor=south west, inner sep=1.5pt] at (sR.north east) {$-\theta$};
  \end{scope}

  \node[lbl, align=center, draw=black!35, rounded corners=2pt, inner sep=4pt, fill=white]
    at (-0.75,3.05) {identify\\[1pt] $z_{2}=T-z_{1}$\\[1pt] $\theta\mapsto-\theta$\\[1pt]
                     {\scriptsize(antipodal point)}};
\end{scope}
\end{tikzpicture}
\caption{The gluing construction. (a) The slice $\Sigma$ with the two generic points $p_{1}$, $p_{2}$ on the manifold and the
balls of radii $R<R_{0}$ around them. (b) After the conformal blow-up each punctured ball
$B_{R}(p_{j})\setminus\{p_{j}\}$ becomes a half-infinite asymptotically cylindrical end with coordinate
$z_{j}=\log(R/r_{j})\in(0,\infty)$ and cross-section $S^{d-1}$, so that $z_{j}=0$ at $r_{j}=R$ and the
point $p_{j}$ is at $z_{j}=\infty$. The tubes are cut at $z_{j}=T$ (the grey parts are discarded) and
the remaining pieces are identified by $(z_{1},\theta)\mapsto(T-z_{1},-\theta)$, where $\theta\in S^{d-1}$
is the unit vector of the Riemann normal coordinates and $-\theta$ is its antipodal point. Thus the starred
point $\theta$ on $z_{2}=0$ is identified with the starred point $-\theta$, on the far side of the other
cylinder, at $z_{1}=T$. The middle of the resulting neck is $z_{1}=z_{2}=T/2$ (red). For simplicity the picture is drawn
for $d=2$, where the cross-section is a circle, however in our construction $d\geq 3$.}
\label{fig3}
\end{figure}

The inclusion of matter fields will proceed as follows; the starting point is as mentioned before, to find a conformal mapping so that the constraint equation satisfies \eqref{mattcons}. Once that is found, we then apply the Weyl transformation for the matter fields corresponding to the mapping, \eqref{balltocylinder}, from the balls to the cylinders where the centers of the balls are mapped to the asymptotic infinity of the cylinders. Then similar to the identification done for the metric, one identifies the the field configurations on a finite segments of the cylinders using the same partitions of unity in the wormhole region, while leaving regions outside the wormhole (and the buffer) unchanged. However, this new field configuration does not necessarily satisfy the matter constraint equations with respect to the glued metric $\tilde{g}_{T}$. In addition the glued metric $\tilde{g}_{T}$ and symmetric traceless tensor $\sigma_{T}$ do not necessarily satisfy the Hamiltonian and momentum constraint with respect to the new glued matter fields. This is achieved, \cite{Isenberg:2005xp}, by first modifying slightly the matter field configuration in the glued region so that the constraint equation is satisfied. This generally involves solving for $\Phi_{T,per}$ in,
\begin{equation}
    C(\Phi_{T,glued} + \Phi_{T,per}, \tilde{g}_{T})=0.
\end{equation}
The next step will be to find $\mu_{T}$, the correction to $\sigma_{T}$ that solves the momentum constraint. Recall that now this constraint does not require the tensor to have zero divergence since it now involves $J_{k}$ coming from the matter fields. The requirement for $\mu_{T}$ as small perturbation will then impose some condition on matter fields, where the goal, which is to find an initial condition that is arbitrarily close to the one we started with in regions outside the wormhole region, is necessarily follows. However, the solution to the constraint equations is not the set, ($\tilde{g}_{T},\sigma_{T}-\mu_{T},\Phi_{T,glued} + \Phi_{T,per}$), rather it should be in the Weyl orbit of this set. Therefore one needs to solve the Licnerowicz equation and find the appropriate $\phi_{T}$ leading to the physical initial condition. This and the conditions that should be satisfied by the matter fields so that the Licnerowicz equation\footnote{One of the main issue is the invertability of linearization of the the Licnerowicz operator. This is possible under certain conditions for the terms in the Licnerowicz equation.} behaves well is discussed in \cite{Isenberg:2005xp}, see also \cite{Witten:2022xxp} for related discussions on the properties of the Licnerowicz equation in the presence of matter and compact extra dimensions. It was also shown those conditions are satisfied for Maxwell and Yang-Mills theories among others where matter constraint equations arise. 

\subsection{Some comments on the wormhole solution}\label{usecom}

As we have stated before the wormhole handle is topologically a cylinder, $I\cross S^{d-1}$. Since we have constructed it by identifying points in each of the asymptotically cylindrical ends, we can adapt the coordinates of either ends to describe it. In particular, at $z_{1}=0$ is a sphere which corresponds to the first mouth of the wormhole while at $z_{1}=T$ is the second mouth beyond $z_{1}=[0,T]$ is the rest of the ambient slice to which the handle is attached. We can similarly say the same in terms of the coordinates of the second end of the original manifold. However, it is more convenient if we define a new coordinate $l$ for the `length' of the cylinder to explicitly state the metric that resulted from the IMP gluing. In particular, let
\begin{equation}
    l=z_{1} -T/2 =-z_{2} +T/2
\end{equation}
where the handle is now given by the segment $l=[-T/2,T/2]$ with $S^{d-1}$ at each point. Then, in this coordinate, the resulting metric from the above construction for the handle is,
\begin{align}\label{handlemet}
    g_{T}&= \phi^{-q}_{T}\tilde{g}_{T}, \text{ and }\\
    \tilde{g}_{T}= dl^{2}&+d\Omega_{d-1} + O(e^{-T/2}\text{cosh}(l)), \text{ while } \phi_{T}= 2R^{-2/q}\;e^{T/q}\text{ cosh}(2l/q)+ \cdots
\end{align}

Of course the first thing to note is that for $l$ close to the mouths $\sim \pm T/2$, the correction to the standard round metric $d\Omega_{d-1}$ becomes relatively large. This can be managed however by looking at the division of the handle into $E_{0,1,2}$ that we have mentioned before. In the regions $E_{1,2}$ the metric on the tubes is glued such that $\tilde{g}_{T}$ is the same as $\tilde{g}_{1,2}$. That is we know what the metric looks like in these regions and they are the original metrics, up to a correction coming from $\phi_{T}$. While on $E_{0}$, it is a linear combination of the metrics on each tube with coefficients that sum to one. If we define $E_{0}$ to be the region given by $l=[-a,a]$ where $a \sim O(1)$ in the large $T$ limit, then the region $E_{0}$ will be given by the metric on a cylinder up to exponentially small corrections in $T/2$. Thus the suggestive way we wrote $\tilde{g}_{T}$ will be reasonable in this region. Coming back to the scale factor, the cosh$(2l/q)$ term comes from the gluing protocol set by IMP, that on most of the neck, i.e, $|l|<l_{0}\sim O(T/2)$, we have $\phi_{T}=\phi_{1}+\phi_{2}$. Since $r_{1}\propto e^{l}$ and $r_{2}\propto e^{-l}$, one arrives at the cosine hyperbolic expression above. The dots again correspond to exponentially small corrections to $\phi_{T}^{-q}$. Combining these two terms, the full metric to leading order in $T$, particularly in the region $E_{0}$, is given by
\begin{equation}\label{gT}
    g_{T}=2^{4/d-2}R^{2}e^{-T} \;\text{ cosh}^{4/d-2}(l(d-2)/2)\;\big(dl^{2}+d\Omega_{d-1}\big) + \cdots
\end{equation}
the dots denoting small corrections.

Even though the above metric does not exactly look like the Einstein Rosen metric for the Schwarzschild \cite{Einstein:1935tc}, it is reasonable to suspect that it should be diffeomorphic to the Schwarzschild metric (here we are interested in Schwarzschild metric in AdS). To check this let's first consider a metric on a cylinder ($\gamma = dl^{2}+d\Omega_{d-1}$) and ask what conformal factor would make it a solution. This question is answered, following our discussion above, by a conformal factor that satisfies the Licnerowicz equation (with no matter) for the cylinder metric. This is given by, check appendix \ref{confmeth},
\begin{equation}
 \Delta_{\gamma}\varphi +\alpha \;\Bigl(R(\gamma)\varphi - \; |\sigma|^{2}_{\gamma}\varphi^{\frac{2-3d}{d-2}}\Bigl) -\beta \varphi^{\frac{d+2}{d-2}} =0, 
\end{equation}
where $\alpha = - \frac{d-2}{4(d-1)}$ and $\beta = \alpha\Lambda $. To simplify our computations we have taken the slice to be a slice with $\tau=0$ (this will be the slice that interests us anyways). In general, $\beta$ will have extra contribution proportional to $\tau^{2}$. 

For a simple metric like $\gamma$, the Licnerowicz equation simplifies significantly. For instance, we have $R(\gamma)= (d-1)(d-2)$ and $\Delta_{\gamma}\varphi = \partial^{2}\varphi/\partial l^{2}$, for $\varphi$ that only depends on $l$. In addition, we want to compare the metric $\varphi^{4/d-2}\gamma$  with the spacelike slice at $t=0$ of the Schwarzchild-AdS metric and thus we have $\sigma=0$ and $\Lambda=-d(d-1)/2l^{2}_{AdS}$. Thus the Licnerowicz equation reduces to,
\begin{equation}\label{lic}
    \varphi^{''} -\frac{(d-2)^{2}}{4} \varphi - \frac{d(d-2)}{8 \;l^{2}_{AdS}} \varphi^{\frac{d+2}{d-2}}=0.
\end{equation}

Note however that if we were to consider a metric of the form,
\begin{equation}
    \frac{d\rho^{2}}{f(\rho)} + \rho^{2}d\Omega_{d-1} 
\end{equation}
for some function $f(\rho)$, we can take the change of coordinates, $\rho = \varphi^{2/d-2}$ and $d\rho/\sqrt{f(\rho)} = \varphi^{2/d-2}dl$ to bring it into the form $\varphi^{4/d-2}\gamma$. That is,
\begin{equation}
    \rho=\varphi^{2/d-2} \text{ and, } \frac{d\rho}{dl}=\sqrt{f(\rho)}\; \varphi^{2/d-2}
\end{equation}
From this we can read of $\varphi$ and $\varphi^{'}$ as functions of $\rho$,
\begin{equation}\label{rhodep}
    \varphi=\rho^{d-2/2} \text{ and, } \varphi^{'}= \frac{d-2}{2}\sqrt{f(\rho)}\; \rho^{d-2/2}
\end{equation}
In addition, from \eqref{lic} we can get a first order differential equation by multiplying it with $\varphi^{'}$ and integrating over $l$, and using the chain rule for $(\varphi^{'}\varphi^{'})^{'}$. Rewriting the first order equation as functions of $\rho$ using \eqref{rhodep}, we get 
\begin{equation}\label{f}
    f(\rho)=1 -\frac{C}{\rho^{d-2}}+\frac{\rho^{2}}{l_{AdS}^{2}}
\end{equation}
where $C$ is the integration constant. This $f(\rho)$ is that of the AdS-Schwarzschild metric for $C=\frac{16\pi GM}{(d-1)\Omega_{d-1}}$. Thus the AdS-Schwarzschild metric is a classical solution  that is (diffeomorphic to a metric) in the Weyl orbit of a metric on a cylinder $\gamma$.  

A slightly puzzling part at this point is that the gluing and the metric $g_{T}$ also do work well for Cauchy slices in flat spacetime. The flat spacetime gluing (for Euclidean and asymptotically hyperbolic slices) was indeed the first version of the IMP gluing before it was generalized to AdS spacetimes too. However, this would make sense if the large $T$ limit of the IMP gluing, $g_{T}$, in AdS spacetime corresponds to the small black hole (white hole) limit of the AdS-Schwarzschild metric. In this limit, the AdS-Schwarzchild black hole metric approaches the standard Schwarzchild metric in flat spacetime. This can be seen from \eqref{f} if we take the limit, $\rho_{h}/l_{AdS}\rightarrow0$, where $\rho_{h}$ is the horizon radius (i.e, $f(\rho_{h})=0$) and is kept fixed in the limit. We find that $\rho_{h}^{d-2}\approx C$ and 
\begin{equation}
    f(\rho)\approx 1-\frac{\rho_{h}^{d-2}}{\rho^{d-2}}.
\end{equation}

 On the other hand, for large $T$, the metric \eqref{gT} can be rewritten with a simple change of coordinates, $r=Re^{-T/2}e^{l}$, as follows,
\begin{equation}
    g_{T} \approx \bigg(1+\big(\frac{Re^{-T/2}}{r}\big)^{d-2}\bigg)^{4/d-2}(dr^{2}+r^{2}d\Omega_{d-1}),
\end{equation}
and this is what is called the isotropic version of the Schwarzchild metric, see for example \cite{Dennison:2010wd}. This is in particular true if we identify,
\begin{equation}
    C=4\;(Re^{-T/2})^{d-2}.
\end{equation}
Thus we find that $\rho_{h}^{d-2}\sim GM \propto (Re^{-T/2})^{d-2}$. The Hawking temperature for the AdS-Schwarzschild is in general given in terms of the horizon radius as,
\begin{equation}
    T_{H}=\frac{1}{4\pi \rho_{h}}\big((d-2)+d\frac{\rho_{h}^{2}}{l_{AdS}^{2}}\big).
\end{equation}
For the IMP glued wormhole the temperature of the each of the mouths will be 
\begin{equation}
    T_{H}\approx\frac{(d-2)e^{T}}{\pi R}.
\end{equation}

It is known that such black holes are thermodynamically unstable and quickly evaporate, and as such most of our discussion in what follows is concerning the $\tau=0$ slice they are constructed on and not about the dynamics.

\subsection{The quantum constraint equations}

In AdS/CFT, quantum gravitational states are given by the states of the boundary theory. As with any QFT, states in the boundary CFT are constructed on a given Cauchy slice, $\partial\Sigma$ at $t=t_{0}$ by a Lorentzian path integral, for instance, with the appropriate initial conditions at infinity in the past and operator insertions. The wave functions are then functionals of the dynamical degrees of freedom on the Cauchy slice. The states we are interested in are for example the thermofield double state at some temperature in the tensor product Hilbert space of two CFTs, or some particular CFT state whose bulk semiclassical interpretation would be a state on a background where a wormhole is connecting different regions of the spacetime, in other words the background have a non trivial $\pi_{1}(\Sigma)$.

The duality dictates that the bulk picture of this story involves a bulk gravitation path integral, formally over all manifolds and metrics\footnote{Here one chooses only allowable metrics, where a matter fields can be consistently coupled in perturbation theory. This path integral is not very well understood non perturbatively. One proposal is to do the integral over the appropriate middle dimensional contour in the space of complex metrics, where the path integral can be seen to converge \cite{Witten:2021nzp}. At perturbative level, however it is better understood and it will be discussed in detail in the following sections.} on the manifolds with a future boundary being a bulk Cauchy slice $\Sigma$ with the boundary Cauchy slice $\partial\Sigma$ as its boundary. It is then expected that with the appropriate sources at the boundary and initial conditions at the far past, one can create states that are functionals of the dynamical degrees of freedom at the bulk Cauchy slice. There are a few things that should be mentioned at this point. The first is that there is no canonical choice for this bulk Cauchy slice, that is there are several bulk Cauchy slices that are related to each other by a small bulk diffeomorphism\footnote{As opposed to large diffeomorphisms, that do not die off at infinity.} and have the same boundary. Therefore, naively, at least at semi classical level one should associate the boundary state to a bulk state defined on the region described in the beginning of the previous section, figure \ref{fig1}, called the Wheeler DeWitt patch. The second point is a state constructed in such a way satisfies what are called the Wheeler DeWitt equations, the quantum versions of the Einstein constraint equations,\cite{Hartle:1983ai,PhysRevD.43.1170,BARVINSKY1993237,Feng:2017xsh}. It is thus prudent to understand these equations to which the bulk state we are interested in is a solution.


The Wheeler DeWitt equations naturally arise in the canonical quantization of the phase space of gravity. In particular the direct canonical quantization of the Wheeler DeWitt patch, which we mentioned in the previous section, with boundary $\partial\Sigma$. Then the dynamical degrees of freedom are the metric $g_{ij}$ on a given Cauchy slice $\Sigma$ in the patch, and the conjugate variable
\begin{equation}
    \Pi^{ij} = \frac{1}{8\pi G} \sqrt{\text{det}g} \bigg(K^{ij} - g^{ij}\text{tr}_{g}K\bigg).
\end{equation}
Upon quantization these are promoted to operators and we have,
\begin{equation}
    \Pi^{ij} = \frac{\delta}{i\;\delta g_{ij}}\;
\end{equation}
Thus the classical constraint equations \eqref{fullconequ} will have a quantum mechanical analogue where the extrinsic curvature is now replaced by derivative with respect to the metric. Therefore the momentum constraint becomes a first order differential operator, while the Hamiltonian constraints is a second order differential operators in the metric. Note that we have tr$_{g}\Pi = g_{ij}\Pi^{ij} =\frac{1}{8\pi G} \sqrt{\text{det}g}\; (1-d)$ tr$_{g}K$.

As previously alluded, these equations act on wave functionals in the Wheeler DeWitt patch and can be roughly understood as imposing diffeomorphism invariance in the patch. In particular, the momentum constraint will impose the condition that the wave functions is invariant under the diffeomorphisms (which are connected to the identity) of the bulk Cauchy slice $\Sigma$. While on the other hand, one can understand the quantum Hamiltonian constraint to corresponds to diffeomorphisms that are normal to the Cauchy slice and relate different Cauchy slices with the same boundary slice, $\partial\Sigma$, see figure \ref{fig1}. These equations are infinitely many partial differential equations for each point on the Cauchy slice, $P_{k}(x)\psi(g)= H(x)\psi(g)=0$.

\providecolor{mydarkblue}{RGB}{20,80,140}  
\providecolor{myblue}{RGB}{50,120,200}
\providecolor{myred}{RGB}{190,30,45}

\begin{figure}[t]
\centering
\begin{tikzpicture}[x=1cm,y=1cm,line cap=round,line join=round,
    declare function={
      a = 2.6;                                   
      sigA(\x)  = 0.32*sin(180*\x/a);            
      sigB(\x)  = 1.05*(1-(\x/a)^2);             
    },
    lbl/.style={font=\small}]

  \fill[black!4] (-a,-4.1) rectangle (a,4.1);
  \foreach \s in {-1,1}{
    \draw[line width=1.0pt] (\s*a,-3.6) -- (\s*a,3.6);
    \draw[line width=1.0pt, dash pattern=on 1.5pt off 2.5pt]
      (\s*a,3.6) -- (\s*a,4.1) (\s*a,-3.6) -- (\s*a,-4.1);}

  \path[fill=mydarkblue!16] (-a,0) -- (0,a) -- (a,0) -- (0,-a) -- cycle;
  \draw[densely dashed, line width=0.6pt, black!65]
    (-a,0) -- (0,a) -- (a,0) -- (0,-a) -- cycle;

  \foreach \x in {-1.95,-1.3,-0.65,0,0.65,1.3,1.95}
    \draw[-stealth, line width=0.5pt, black!60]
      (\x,{sigA(\x)+0.05}) -- (\x,{sigB(\x)-0.06});

  \draw[line width=1.1pt, myblue, domain=-a:a, samples=80, smooth]
    plot (\x,{sigA(\x)});
  \draw[line width=1.1pt, myred, domain=-a:a, samples=80, smooth]
    plot (\x,{sigB(\x)});

  \foreach \s/\anc in {-1/east,1/west}{
    \fill (\s*a,0) circle[radius=2pt];
    \node[lbl, anchor=\anc, inner sep=4pt] at (\s*a,0) {$\partial\Sigma$};}

  \node[lbl, myred]  at (0,1.32)      {$\Sigma_{0}$};
  \node[lbl, myblue] at (1.35,-0.62)  {$\Sigma$};
  \node[lbl, black!70] at (0.33,0.45) {$u$};
  \node[lbl, mydarkblue] at (0,-1.55) {WdW patch};
  \node[lbl, mydarkblue] at (1.95,3.45) {$\mathcal{M}$};
  \node[lbl, rotate=90, black!70] at (-a-0.28,2.6) {$\partial\mathcal{M}$};
  \node[lbl, rotate=-90, black!70] at (a+0.28,2.6) {$\partial\mathcal{M}$};
  \draw[-stealth, line width=0.5pt, black!70] (-a-1.25,-3.4) -- ++(0,1.2)
    node[lbl, above, text=black!70] {$t$};
\end{tikzpicture}
\caption{The Wheeler--DeWitt patch (shaded) associated with a boundary time slice $\partial\Sigma$ is the
set of bulk points of $\mathcal{M}$ that are spacelike separated from $\partial\Sigma$; its boundary
(dashed) is null. It is the domain of dependence of any Cauchy slice $\Sigma$ anchored at $\partial\Sigma$.
A different Cauchy slice $\Sigma_{0}$ in the patch, anchored at the same $\partial\Sigma$, is related to
$\Sigma$ by a small diffeomorphism, that is, a diffeomorphism along a vector field $u$ that vanishes at
$\partial\Sigma$. In this sense the two Cauchy slices are equivalent. In our context, we consider the closure of the Wheeler deWitt patch by including the spacelike boundary $\partial\Sigma$.}
\label{fig1}
\end{figure}
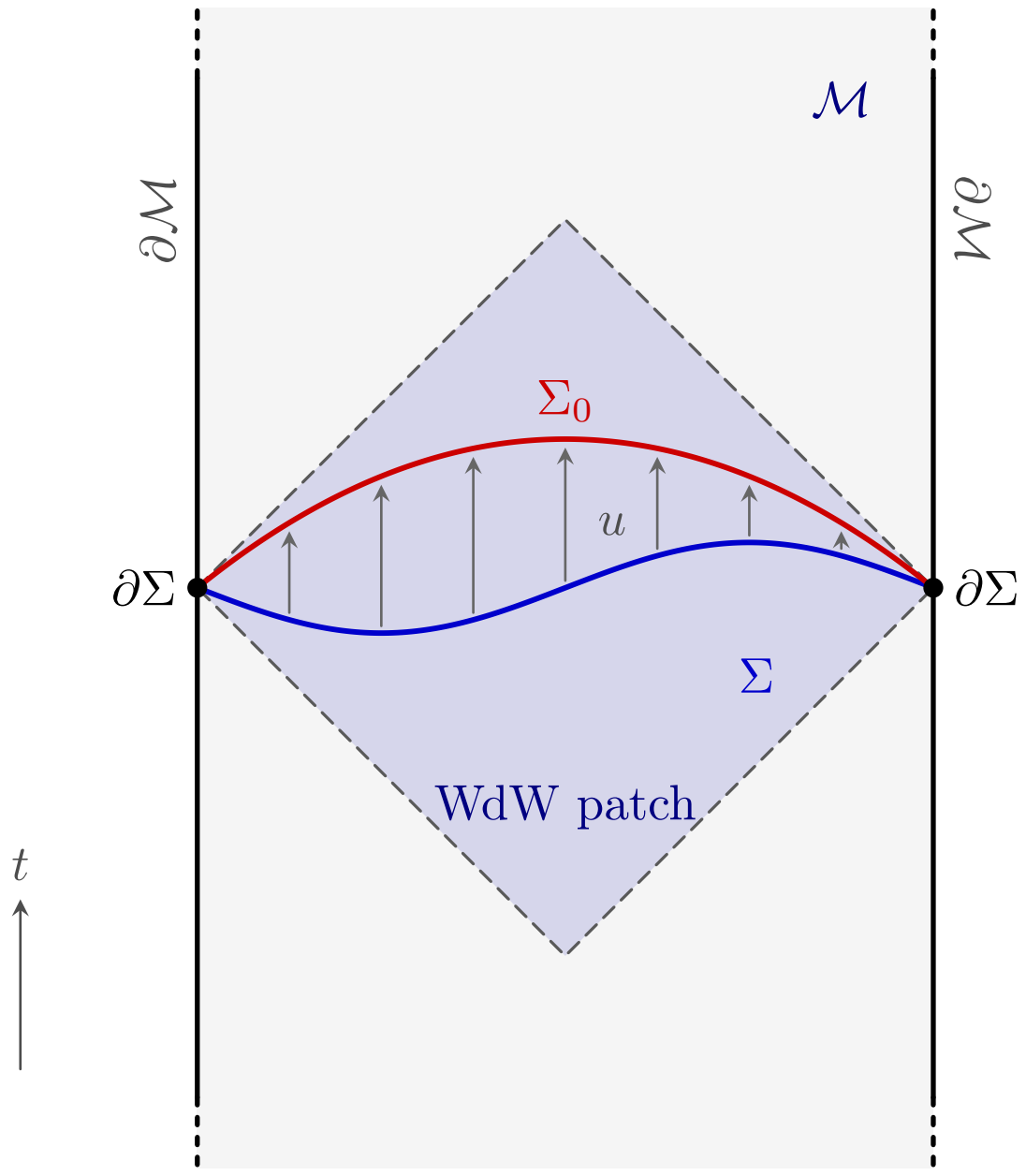

One can also consider the dynamical degrees of freedom discussed in the previous section, which are the conformal class of the metrics, $[g_{ij}]$, and the trace of the extrinsic curvature, $\tau$. Then the conjugate variables would be the traceless part of $\Pi^{ij}$ and volume measure $\text{det}g$ of $\Sigma$. Now the classical constraint equation will imply a different looking quantum constraint equations but still two equations, one first order and the other second order differential equations which in general are equally involved as the previous ones. The states $\psi([g_{ij}],\tau)$ were considered in \cite{York:1972sj} in the hopes that the trace of the extrinsic curvature provides the time coordinate in quantum gravity. For us however, this problem of introducing a time variable does not arise since time is already well defined at the boundary as the boundary CFT time coordinate. These states and the constraints can be seen to be well suited however for a discussion in the context of BRST quantization introduced in \cite{Witten:2022xxp} which we quickly review in the following subsection.     

\subsection{`BRST quantization' \`a la Witten}\label{boundgaufix}

Since the Wheeler DeWitt equations are imposing gauge invariance condition on the states, it would be expected that a refined framework for discussing these equations is BRST quantization. This is of course not the full BRST quantization of the theory but only of the slice the states are being prepared on. There is additional gauge fixing needed for the path integration to the past and future of the slice. In other words, the particular BRST treatment that would impose the Wheeler DeWitt equations is the one that acts on the Cauchy slice $\Sigma$ where the dynamical coordinates live.

The ghost fields will thus be vector fields defined on $\Sigma$, $c^{\mu}(x)$, with the opposite statistics and the action the BRST transformation on the ghosts, the metric and the matter fields will be just like a diffeomorphism transformation but with a parameter $c^{\mu}$. The antighost multiplet as usual will transform as 
\begin{equation}
    \delta_{Q}(\bar{c}^{\mu}(x)) = b^{\mu}(x), \; \text{ and } \delta_{Q}(b^{\mu}(x)) =0 
\end{equation}
where $b^{\mu}$ is a bosonic vector field with ghost number zero, while for the $c^{\mu}$ and $\bar{c}^{\mu}$ it is one and minus one respectively. 
The BRST charge in general will look like,
\begin{equation}\label{Q}
    Q = \int_{\Sigma} d^{d}x \sqrt{\text{det }g}\; \bigg( c^{0}(x) H(x) + c^{k}(x)P_{k}(x)+(\text{higher order terms in the ghosts})\bigg)
\end{equation}
The BRST charge satisfies $Q^{2} = \delta_{Q}^{2}=0$ and the physical space of states satisfying the Wheeler deWitt equations is defined as the cohomology class, $\mathcal{H}_{\text{phy}} = \frac{\text{ker }Q}{\text{im }Q}$, of the BRST charge operator. Thus one has several ways of imposing the quantum constraints depending on the choice of the ghost number sector for the physical space of states. For instance choosing all the anti ghost modes to annihilate the states will imply that $P_{k}(x)\psi_{\text{phy}}(g)= H(x)\psi_{\text{phy}}(g)=0$. On the other hand, choosing $c^{k}$ and $\bar{c}^{0}$ to annihilate the states of $\mathcal{H}_{\text{phy}}$, following the commutation relation of the $c$ and $\bar{c}$'s, implies that $H(x)\psi_{\text{phy}}(g)=0$, while the equivalence relation $\psi_{\text{phy}}(g) \simeq \psi_{\text{phy}}(g) + Q \chi$ for any state $\chi$ 'replaces' the momentum constraint. And similarly for other choices of the ghost and antighost fields annihilating physical states. 

The BRST quantization discussed in \cite{Witten:2022xxp} and which we will make use of here is a slightly different one. It makes use of the the conformal degrees of freedom we have been discussing so far, i.e, the dynamical coordinates are the conformal class of the metric $[g]$ and the mean curvature $\tau$ rather than the full metric on $\Sigma$. The quantum constraint equations on the states $\psi([g],\tau)$ involving these coordinates are as complicated as the ones on $\psi(g)$, involving the naive dynamical coordinates. However, the conformal coordinates makes manifest a certain gauge fixing that simplifies our discussion. This is a gauge fixing that sets $\tau = \text{tr}_{g}K =0$ and thus fixes $\Sigma$ to be a volume maximizing slice\footnote{If we let $t$ be a time coordinate that parametrizes the geodesic distance in the direction $n^{\mu}$, normal vector to $\Sigma$, and let $g(x) = g(x,t=0)$. The condition that $\Sigma$ is volume maximizing is the statement that the variation of the volume vanishes, $\delta_{\epsilon}V=\int_{\Sigma}d^{d}x\; \sqrt{\text{det }g(x,t=\epsilon)} - \sqrt{\text{det }g(x,t=0)} = 0$. It is then easy to show that $\delta_{\epsilon}V = \int_{\Sigma}d^{d}x\; \sqrt{\text{det }g(x)} \; \epsilon\tau + O(\epsilon^{2}).$ For asymptotically AdS and AdS$\cross Y$ spaces where $Y$ is a compact manifold, the volume of a spacelike slice is infinite and has to be renormalized, so what one really extremizing is the renormalized volume.}. Now the states look like $\psi([g],\tau=0)$ and are prepared on a unique volume maximizing slice $\Sigma$, which we will discuss in more detail in the next section. We want to also note that this gauge fixing is convenient for the conformal wormhole surgery of the underlying background spacetime since it is applies for constant mean curvature slices and simplifies when $\tau=0$. In any case, To impose this gauge fixing, we add the following term to the action in the path integral,
\begin{equation}\label{slicegauge}
    \delta_{Q}\int_{\Sigma} d^{d}x \sqrt{\text{det }g}\; (\bar{c}^{0}\tau),
\end{equation}
to impose the gauge fixing. This will eventually lead to $c^{0}=0$ as an operator, since the variation $\delta\tau$ due to an infinitesimal diffeomorphism, $\epsilon$, that modifies the slice along some time like direction, is itself proportional to $\epsilon$. Thus the BRST transformation $\delta_{Q}\tau$ will be proportional to $c^{0}$. On the other hand the condition $\tau=0$ fixes $\Sigma$ to be the unique volume maximizing slice, therefore the Hamiltonian constraint will be absent in this gauge choice. There is still, however, the momentum constraint for which we proceed just as above. Then, for instance, after choosing the $c^{k}$'s to annihilate the physical states one can see that the BRST equivalence relation, $\psi_{\text{phy}}([g],\tau=0) \simeq \psi_{\text{phy}}([g],\tau=0) + Q \chi$, implies that states like $P_{k}(x)\chi$ are identified with null vectors. This enforces the diffeomorphism invariance on $\Sigma$ for the physical states. Because of this gauge fixing, states prepared in such a way are associated to the specific volume maximizing slice and not to the full Wheeler DeWitt patch.

\section{The quantum states}\label{states}

\subsection{Path integral preparation of the states}\label{pathintstat}

The BRST quantization discussed in the previous section can be considered as imposing gauge invariance on the future boundary of the path integral used to prepare the ket states (or past boundary of the path integral that we use to prepare the bra states), that is $\Sigma$. Therefore, it's still left to impose the gauge constraints in the spacetime region to the past (or future) of the Cauchy slice $\Sigma$. Recall that we are interested in perturbation theory, in particular we are want to consider states $\psi([g],\tau=0)$ for metrics $g$ sufficiently close to some classical metric $g_{0}$, in other words $g=g_{0}+h$, where $h$ is a perturbation around $g_{0}$. It is clear that as far as these states are concerned, $g_{0}+h\simeq g_{0}$ if $h$ is proportional to $g_{0}$. 

To prepare such states, we need to do a path integral over metrics (with the appropriate boundary operator insertions) that are again close to some classical solution which is determined by the appropriate classical initial data on $\Sigma$. Before discussing this, we first consider the preparation of the states with the more familiar non-elliptic boundary condition. In this case, the states, $\psi(g)$, are functionals over the space of metrics on $\Sigma$ and satisfy the quantum constraint equations. The metric $g$ is expanded around a classical solution, $g=g_{0}+h$, and the path integral is over spacetime metrics, $\hat{g}_{\mu\nu}$, on a manifold, $\mathcal{M}_{+}$, with $\Sigma$ as its future boundary. As we mentioned before the spacetime metrics on $\mathcal{M}_{+}$ are also close to a classical solution, i.e, $\hat{g}_{\mu\nu}=\bar{g}_{\mu\nu} + \hat{h}_{\mu\nu}$, where $\bar{g}_{\mu\nu}$ is a classical solution with $\bar{g}_{ij}|_{\Sigma}=g_{0,ij}$. We do not fix the rest of the components of $\bar{g}_{\mu\nu}|_{\Sigma}$; on the other hand, for the perturbations we have  $\hat{h}_{ij}|_{\Sigma} = h_{ij}$ and the restriction of the de Donder gauge to $\Sigma$ (discussed in the next subsection) fixes the rest of the components of $\hat{h}_{\mu\nu}$. As we review in appendix \ref{b}, these boundary conditions will lead to serious problems in perturbation theory.

Note that choosing the metric $g_{0,ij}$ and the extrinsic curvature $K_{0,ij}$ on the slice $\Sigma$ fully determines the classical solution on $\mathcal{M}_{+}$, which we called $\bar{g}_{\mu\nu}$. To compute $\psi(g_{0})$, we do the path integral on $\mathcal{M}_{+}$, by considering metric perturbations to $\bar{g}_{\mu\nu}$ that vanish on $\Sigma$. Therefore, as we do the path integration, $K_{0,ij}$ will not be kept fixed, only $g_{0}$ will. Computing the wave functional in the presence of perturbation like $g_{0,ij}+h_{ij}$ on $\Sigma$ will proceed as described above. Let's now come back to the states constructed by a better behaved boundary conditions, $\psi([g_{0}+h],\tau=0)$. Here again $g_{0,ij}$ and $K_{0,ij}$ determine a classical solution $\bar{g}_{\mu\nu}$ on $\mathcal{M}_{+}$, however the path integral is done with perturbations to $\bar{g}_{\mu\nu}$ that preserve the conformal structure of the metric on $\Sigma$ and the mean curvature $\tau= g_{0,ij}K^{ij}_{0}$. Thus, when we do the path integral, the conjugate variables which are the scale factor of the metric and the traceless part of $K_{0}$ are not kept fixed and do fluctuate. More precisely, we can perform the path integral over metrics, $\hat{g}_{\mu\nu}=\bar{g}_{\mu\nu} + \hat{h}_{\mu\nu}$, with $\hat{h}_{ij}|_{\Sigma}=\gamma \;\bar{g}_{ij}|_{\Sigma}$. In addition the perturbations are such that, 
\begin{equation}
    \hat{g}_{ij}\hat{K}^{ij}|_{\Sigma}=\tau,
\end{equation}
where $\hat{K}_{\mu\nu}$ is the extrinsic curvature as a function of the full metric $\hat{g}$. This path integral computes the value of the wave functional when the conformal class of the metric is given by $[g_{0}]$, on the Cauchy slice $\Sigma$ with mean curvature $\tau$ (which in our context equals zero). To compute the wave functional in the presence of perturbations, $h_{ij}$, that modify the conformal class of $g_{0}$ to $[g_{0}+h]$, we do the path integral over perturbations with boundary conditions
\begin{equation}
   \hat{h}_{ij}|_{\Sigma}=\gamma \;\bar{g}_{ij}|_{\Sigma} +(1+\gamma)\;h_{ij}
   \end{equation}
so that $$\hat{g}_{ij}|_{\Sigma}=(\bar{g}_{ij}+\hat{h}_{ij})|_{\Sigma} \in [g_{0}+h].$$ In addition the perturbations have to satisfy $ \hat{g}_{ij}\hat{K}^{ij}|_{\Sigma}=\tau$. 

The above path integral projects the ket states onto the boundary conditions on $\Sigma$. To prepare the bra states, one follows the same procedure but now on the manifold $\mathcal{M}_{-}$ where $\Sigma$ is its past boundary. the path integral then prepares the complex conjugate wave functional, $\psi^{\star}([g],\tau=0)$.

\subsection{More on the states}

In summary, given some $D$ dimensional asymptotically AdS manifold, $\mathcal{M}$, and a classical solution of the Einstein field equations, $\bar{g}$, we would like to consider perturbative states in $G_{N} \rightarrow 0 $ limit around this classical solution. We want these perturbative states to be dual to states of the boundary CFT, defined on some boundary Cauchy slice $\partial\Sigma$. To this end, we consider a Cauchy slice of the classical solution $\Sigma \subset \mathcal{M}$ with boundary $\partial\Sigma$ and mean curvature $\tau = 0$. For asymptotically AdS$_{D}$ spacetimes, such a slice is one of the slices in the Wheeler- DeWitt patch. If we also include the boundary of the patch, $\partial\Sigma$, so that the Wheeler-deWitt patch becomes compact, it can be rigorously shown that, $\Sigma$ (a slice with $\tau=0$ and boundary $\partial\Sigma$) exists and is unique \cite{bonsante2009maximalsurfacesuniversalteichmuller,Chrusciel:2022cjz}. For asymptotically AdS$\cross Y$ spacetimes, where $Y$ is some compact space, there are physical arguments for the existence and uniqueness of such a slice but rigorous proofs are still missing \cite{Witten:2022xxp}.

The existence and uniqueness condition is an elliptic condition and therefore is an open condition. This means a small deformation of a classical metric where such a $\Sigma$ exists will still be a metric where such a slice exists. Thus we can do the path integration to construct the perturbative states on the slice with $\tau = 0$ of the modified metric $\hat{g}$, to all orders in perturbation theory. More precisely, this condition is implemented by the addition of the term \eqref{slicegauge} to the action in the path integral. Because of this BRST condition, we can not fix the full metric on the Cauchy slice, we can only fix the conformal class of the modified metric. Thus the states we consider are functionals of the conformal class $[g]$ of metric on $\Sigma$ with a perturbative expansion around the classical solution $g_{0}$ to all orders in $G_{N}$, defined on a slice with $\tau=0$ that is bounded by $\partial\Sigma$.

Note that we have not fully imposed gauge fixing for the path integral preparing $\psi([g], \tau =0)$ yet. We do this following the standard BRST quantization of the theory with for example the de Donder gauge. For completeness, we summarize the quantization below. 

The BRST transformation of the metric and the matters fields will be the same as their diffeomorphism transformation but now the ghost fields $\hat{c}^{\mu}$ replacing the gauge parameters, i.e,
\begin{align}
    \delta \hat{h}_{\mu\nu} &= D_{\mu}\hat{c}_{\nu} + D_{\nu}\hat{c}_{\mu}\\
    \delta\hat{c}_{\mu} &= \hat{c}_{\nu} \partial^{\nu}\hat{c}_{\mu}.
\end{align}
where $\hat{c}_{\mu}$ transforms just like any other vector field, and $D_{\mu}$ is the covariant derivative with respect to the classical background metric. The antighost multiplet on the other hand transform as usual,
\begin{align}
    \delta\hat{b}_{\mu} &= f_{\mu}\\
    \delta f_{\mu}&=0
\end{align}
where $\hat{b}_{\mu}$ is the antighost field and $f_{\mu}$ the auxiliary field. The gauge fixing action will be,
\begin{equation}
    S_{gf} \propto \delta \int_{\mathcal{M}_{+}} d^{D}x \sqrt{\text{det }\hat{g}}\bigg( -\frac{1}{2}\hat{b}_{\mu}f^{\mu} + \hat{b}_{\mu}T^{\mu}(\hat{h})\bigg)  
\end{equation}
where $T_{\mu}(\hat{h}) = 0$ is the de Donder gauge condition, $\mathcal{M}_{+}$ is the spacetime region to the past of the slice with $\tau = 0$ and boundary $\partial\Sigma$. We set the boundary value of the antighost $\hat{b}_{\mu}|_{\Sigma}$ to zero because we have already imposed a gauge fixing that fixes the mean curvature of $\Sigma$ in section \ref{boundgaufix}. This in turn leads to $\delta\hat{b}_{\mu}|_{\Sigma} =  T_{\mu}|_{\Sigma} = 0$. We will refer to this condition when we discuss the issues resolved by considering conformal dynamical coordinates in appendix \ref{b}. It follows directly from the fact that $\delta\hat{b}_{\mu}=f_{\mu}$ and integrating out $f_{\mu}$. The path integral is done over all metric fluctuations $\hat{h}_{\mu\nu}$ with the appropriate boundary condition that is fixed $[g]$ and $\tau=0$, around some classical solution like we discussed in the previous section. As mentioned before the scale factor of $g$ and the traceless part of the extrinsic curvature on the Cauchy slice fluctuate in these states.

In section \ref{confwormsurg}, we discussed how one starts with an initial data set and modifies it only in some bounded neighborhood of two generically chosen points and adds a wormhole neck. The new initial data set approximates the original initial data set arbitrarily well on the regions away from the handle. An initial data set includes a spacelike manifold, $\Sigma $ and a Riemannian metric  $g_{0}$ and the extrinsic curvature $K = \sigma + \frac{\tau}{d}g_{0}$ (as we stated before we consider a constant mean curvature, in particular $\tau =0$) that solves the Einstein constraint equations. Using the path integral like we discussed above, we can associate to this initial data set a perturbative sets of states (to any perturbative order in $G_{N}$). This would be dual to the large $N$ perturbative boundary quantum states on $\partial\Sigma$. These states however depend on the conformal class of the metric on $\Sigma$ with $\tau = 0$, $\psi([g_{0}+h],\tau)$, where $h$ now encodes all the perturbative corrections to the classical solution of the constraint equations. We now recall the conformal blow up procedure considered in section \ref{confwormsurg}, where we modified the data in two balls each centered around the two points with a radius of geodesic distance $R+\Delta R$. The classical metric, $g_{0}$ is modified by the Weyl factor \eqref{balltocylinder}
\begin{equation}
    \tilde{g}_{0} = \phi^{q}g_{0} \in [g_{0}].
\end{equation}
We also consider a similar blow up for metric on $\Sigma$ that is modified by fluctuation, i.e, 
\begin{equation}
    \tilde{g} = \phi^{q}(g_{0} + h)\in[g_{0}+h].
\end{equation}
we have said that the scale factor of the metric on $\Sigma$ fluctuates, and this is due to the boundary condition for the perturbative corrections for the full metric on $\mathcal{M}_{+}$. This is fine since the wave functional depends only on the conformal class of the metric (and the mean curvature). For this same reason, the conformal blow up procedure will not modify the states.  

In the classical limit, this geometry is indeed the asymptotically cylindrical manifold discussed in subsection \ref{confwormsurg} (the inclusion of the two points which are centers for the balls correspond to including the asymptotic infinity of the cylindrical ends). The latter geometry will be a perturbation of this classical manifold with perturbations that blow up at the cylindrical ends and the same as $g_{0}+h$ outside the ball regions (and the buffer). However, as we have said the quantum state that is going to be associated with each of these geometries is the same one, since one is in the Weyl orbit of the other, $$\psi([g_{0}+h], \tau = 0) = \psi([\tilde{g}],\tau=0).$$
In other words, we have not modified the states since we have not modified the conformal class of the metrics $g_{0}+h$.

On the other hand, we can consider the glued metric that looks like \eqref{partglu} in the handle region and looks like the above metrics in the rest of the region, with some perturbation, i.e, $\tilde{g}_{T} + \tilde{h}_{T}$. In the limit $G_{N} \rightarrow 0$, these metric do not go to a solution of the classical constraint equations, rather they approach $\tilde{g}_{T}$ which does not solve the constraint equations with the similarly constructed $\sigma_{T}$. However one can find a solution of the constraint equations, $g_{T}$ (together with $K_{T}$) that \emph{is} in the Weyl orbit of the $\tilde{g}_{T}$ and $\sigma_{T}-\mu_{T}$, for small $\mu_{T}$. Therefore, for some particular $\phi_{T}$,
\begin{equation}
    g_{T}+ h_{T}= \phi^{-q}_{T}(\tilde{g}_{T}+\tilde{h}_{T})
\end{equation}
where $h_{T}$ is the metric perturbation to the metric $g_{T}$, which is the solution of the constraint equations with the extrinsic curvature $K_{T}$ with $\text{tr}_{g_{T}}K_{T}=0$. This implies, $$\chi([\tilde{g}_{T}+\tilde{h}_{T}],\tau = 0) = \chi([g_{T}+h_{T}],\tau = 0).$$  
However, since $g_{T}$ is a metric that is the same as $g_{0}$ in all but some region (in particular the wormhole region), for any state $\psi([g_{0}+h], \tau = 0)$, there will be some state $\chi([g_{T}+h_{T}],\tau=0)$ in the set of states associated with $\Sigma_{T}$ ($\Sigma$ with a neck topologically added to it) that looks the same in the regions far away from the wormhole region\footnote{This statement becomes precise in the large $T$ limit.}. This will be a central point in the following sections. 

To be more explicit, we consider the $E_{0}$ region of the wormhole handle. The original background metric and the background metric of the wormhole solution away from the handle are all arbitrary smooth slices with constant mean curvature. However, the IMP gluing condition fixes the behavior of the background metric near the horizon, which is given in \eqref{handlemet}. One can compute the expectation value of the metric, $g_{ij}$, in the state $\chi([g_T+h_T],\tau=0)$ in $E_0$. The state $\chi$, constructed by path integration with no operator insertion, is not written in the basis of the metric, it is more convenient to write to pull out the scale factor of the operator as $g_{ij} = \phi^{-q}\mathbf{g}_{ij}$, where $\mathbf{g}_{ij}$ is a representative for the conformal class of the operator. 
This representative can be chosen for example by specifying the determinant, $\text{det}\mathbf{g}$. We choose it to be that of the determinant of the standard metric on a unit sphere in $d-1$ dimensions. The choice is made to match with our earlier and later discussions. As one would expect, the expectation value of the operator $\mathbf{g}_{ij}$ in the `ground state'\footnote{This is not really a ground state in the standard sense as, in general, there is no conserved positive definite quantity to minimize. By the term we just mean that it is prepared by the gravitational path integral without operator insertions to the past of the slice $\Sigma$.}, $\chi$ is highly picked around the conformal class of the cylinder metric in $E_{0}$,
\begin{equation}
    \braket{\mathbf{g}_{T}}|_{E_0} = dl^{2} +d\Omega_{d-1}  + O(e^{-T/2}\text{cosh}(l)) + O(G_{N})
\end{equation}
Coming to the operator $\phi$, the way it acts on $\chi$ is dictated by its dependence on $\mathbf{g}$ and the traceless part of $\Pi_{ij}$, acting as a differentiation operator in terms of $\mathbf{g}$. The reason for this is that, \cite{Witten:2022xxp}, the Licnerowicz equation, which follows from the Einstein's equation of motion for the $00$ component of the metric, relates it to the representative of the conformal class of the metric and the traceless part of the extrinsic curvature. Thus in the path integral, one can do a field redefinition for the $00$ component of the metric and thus replace $\phi$ by a non trivial combination of $\mathbf{g}_{ij}$ and $\frac{\delta}{\delta\mathbf{g}}$. To leading order in $G_{N}$, the $\delta/\delta\mathbf{g}$ term will not be present (as we will see, \eqref{K}) and the IMP gluing will dictate that on $\chi$, we have
\begin{equation}
    \braket{\phi_T^{-q}}|_{E_{0}}= 2^{-q}R^{2}e^{-T}\text{cosh}^{-q}(2l/q)+O(G_{N})+ \cdots 
\end{equation}
localizing on the unique scale factor for the representative $\braket{\mathbf{g}}$. The dots represent exponentially small corrections in $T$ relative to the leading term. Thus $\braket{g_{ij}}\approx \braket{\phi_T^{-q}}\braket{\mathbf{g}_{T}}$ localizes at the solution of the constraint equations on the slice and gives \eqref{handlemet} in $E_{0}$, to leading order in $G_{N}$.

\section{Near and far regions of the wormhole neck}\label{wormcondexp}

Given a classical solution to the Einstein field equations, we constructed above a gravitational perturbative Hilbert space of states, on $\Sigma$ with $\tau=0$ and boundary $\partial\Sigma$, with the conformal coordinates being the dynamical coordinates. The boundary $\partial\Sigma$ is the slice where the CFT dual states are defined on. But since the states on $\Sigma$ depend on the conformal class of the metric on $\Sigma$, we can as well take them to be prepared on a slightly modified background where, two balls centered on two generic points on the original classical solution are blown up to asymptotically cylindrical regions (where we also include the asymptotic ends). This is possible since these ball regions of the original solution can be mapped conformally to (asymptotically ) cylindrical regions. On the other hand, a classical background with a handle where the wormhole mouths are each of the boundaries of the previously mentioned balls, can be conformally mapped to the above background where we mapped the balls to asymptotically cylindrical regions but now points on one cylindrical region are identified with points on the second cylindrical region. This identification is done on points length $T$ away from the boundary of the balls. However, after the identification, the background with the wormhole can be made to be arbitrarily the same as the original background in the region outside the handle by taking $T$ to be big. Thus we are really comparing gravitational perturbation theories on a given classical background with a Cauchy slice $\Sigma$ and a background with a Cauchy slice $\Sigma_{T}$ that we can make as close as we like to $\Sigma$ in the region outside the handle by increasing $T$. Thus whatever new effect arises in the second theory, we expect it to be coming solely from the presence of the wormhole neck. We aim to explore these new effects. We can investigate this by comparing the theories defined on a background with two cylindrical ends and a background where parts of the cylindrical regions are identified.  

\subsection{The QFT on curved spacetime case}

Before discussing the comparison of the two gravitational theories, we first discuss the comparison of quantum field theories on these two backgrounds, one with asymptotically cylindrical regions and the one where the cylindrical regions are identified. We recall that the identification between the two cylinders, each with length $T$, is given by first mapping the points on one of the cylinders, say the second tube region with coordinates $z_{2}$ and $\theta_{2}$ as follows; if $(z_{2},\theta_{2}) = (z,\theta)$ then we identify it with $(z_{1},\theta_{1}) = (T-z,-\theta)$. For a QFT defined on this manifold this identification translates to a certain kind of combination of the von Neumann algebra operators in the cylinders of length $T$ in the limit $T$ is large. In the case that the initial Cauchy slice $\Sigma$ is a disconnected one, we consider a copy of the same theory on each of the two components in the rest of the section. 

Let $\mathcal{A}(N_{i,T})$ be the $*$ algebra of bounded operators generated by the operators in the tube regions of length $T$, where $i=1,2$. If we want to be more precise, $\mathcal{A}(N_{i,T})$ is the algebra of operators associated with the domain of dependence of $N_{i,T}$, and the operators are bounded functions operators that are smeared on a spacelike pancake open neighborhood of $N_{i,T}$ given by a thickening of the spacelike subregions, assuming a causal relativistic quantum field theory. 

The von Neumann algebra of operators associated with $N_{i,T}$ are given by the von Neumann closure of the above * algebras on thickened tube regions to $\hat{N}_{i,T}$. This is expected to be same as the von Neumann algebra of operators in the domain of dependence of the tube regions, $D(N_{i,T})$. For the moment however, we focus on the * algebra of operators on the regions $\hat{N}_{i,T}$. Recalling the identification of the two cylinders mentioned above, if the smeared operator $\phi_{f_{1}}$ is localized around $(z_{1},\theta_{1})$ and the operator $\phi_{f_{2}}$ is localized around $(z_{2},\theta_{2})$ are operators acting in the first and second tube regions respectively before the identification. After the identification, let's call the smeared operator acting in the neck region $ \phi_{f,T}$. We have taken identical test functions $f,f_{1}$ and $f_{2}$ that are centered around distinct points. The IMP gluing of the classical background suggests dividing the handle into three subregions $E_{0,1,2}$; where $E_{1}$ shares a boundary with $E_{0}$ and the rest of the spacetime through the first wormhole mouth, while $E_{2}$ shares a boundary with $E_{0}$ and the rest of the spacetime through the second mouth of the wormhole. the region $E_{0}$ is the intermediate region of the neck sharing boundary with both $E_{1}$ and $E_{2}$. In the region $E_{1}$, we identify the smeared operators therein with the smeared operators in the corresponding $E_{1}$ region of $N_{1,T}$, 
\begin{equation}\label{con1}
    \phi_{f,T}=  \phi_{f_{1}}, 
\end{equation}
when the localization of the smearing functions is in the region $E_{1}$. On the other hand, when the localization of the smearing functions is in the region $E_{2}$, we have
\begin{equation}\label{con2}
    \phi_{f,T}=  \beta(\phi_{f_{2}}), 
\end{equation}
where now if we have a map $\alpha:N_{2,T}\rightarrow N_{1,T}$ takes $(z_{2},\theta_{2})=(z,\theta) \mapsto (z_{1},\theta_{1})=(T-z,-\theta)$, then the map $\beta$ on the smeared operators maps the smearing functions that are localized in $N_{2,T}$ to smearing functions localized in $N_{1,T}$ according to $\alpha$ and it satisfies the expected properties.

\begin{equation}
    \beta(\phi_{f}) = \phi_{\alpha^{*}f}
\end{equation}
where $\alpha^{*}f(x) = f(\alpha(x)) $, and $f$ is a smearing function, and,
\begin{align}
    \beta(AB)&=\beta(A)\beta(B)\\
    \beta(cA) &= c\beta(A)\\
    \beta(A^{\dagger})&=\beta(A)^{\dagger}
\end{align}
for $A,B$ smeared local fields affiliated with $\mathcal{A}(N_{2,T})$ and $c\in\mathbb{C}$.

What we mean by the above identification is really the commutator between operators is the same and the $*$ algebra of operators in the $E_{1}$ region of the wormhole background, $\mathcal{A}(E_{1})_{W,T}$ is identical to the $*$ algebra of operators in the $E_{1}$ region of the original background with two asymptotically cylindrical ends, $\mathcal{A}(E_{1})_{T}$. In addition, as we will see, we will choose states for these algebras the give the same expectation values for the identified operators. The same goes for the $E_{2}$ regions. This is however not the full story and we modify this statement slightly later on.

The full algebra of operators for the quantum field theory defined on the background with the two tubes of length $T$ (after the conformal blow up of the Riemann normal coordinates) is given by, $\mathcal{A}_{T}$ which is given by the bounded functions of the smeared operators, $\phi_{f}$, on an open set containing the spacelike slice. We have $\mathcal{A}(E_{1})_{T}\subset\mathcal{A}(N_{1,T}), \mathcal{A}(E_{2})_{T}\subset \mathcal{A}(N_{2,T})$, and $\mathcal{A}(N_{1,T}),\mathcal{A}(N_{2,T}),\mathcal{A}(M)\subset \mathcal{A}_{T}$ where $M$ is the subregion of the Cauchy slice given by $\Sigma/(B_{1}\cup B_{2})$. It is important to note that the background metric with the conformal blow up $\tilde{g}$ and the corresponding extrinsic curvature do not satisfy the classical Einstein constraint equations. But to define QFT on a curved background, this does not bother us. On the other hand, the full algebra of operators for the QFT on the spacetime where the tube regions are identified (to form a handle) is given by $\mathcal{A}_{W,T}$ where we have $\mathcal{A}(E_{1})_{W,T},\mathcal{A}(E_{2})_{W,T}, \mathcal{A}(E_{0})_{W,T}\subset\mathcal{A}(N_{T})$, and $\mathcal{A}(N_{T}),\mathcal{A}(M) \subset \mathcal{A}_{W,T}$. 

As we mentioned earlier, we will have to modify the statement concerning $\mathcal{A}(M)$ in the Cauchy slice with a wormhole and without a wormhole. The reason what we said above is not completely correct is that, in the cases where we have a gauge field, there is a non zero flux from each of the mouths of the wormhole which are not present in the original theory. The original theory is theory with asymptotically cylindrical ends, which we said are compactified. Consider for instance a $U(1)$ gauge field $A$ and a subregion given by $M\cap B^{0}_{i}$ which is an annulus region for each $i$. We can now consider a local operator $\star F$ integrated over on a $d-1$ oriented cycle that wraps around either of the mouths of the cylinders in the specified region, where $F=dA$ is the field strength. Then the operator $Q_{e}=\oint_{\mathcal{S}_{d-1}}\star F$ which is  a smeared operator is affiliated to the $*$ algebra of operators associated with the region. However, if $M_{d-1}$ is the boundary of $N_{i,T=\infty}$  the operator $Q_{e}$ can be written alternatively using Stoke's theorem, $Q_{e}=\int_{N_{i}}d(\star F)$. This, in the absence of matter gives $Q_{e}=0$. For $d=3$ we can similarly construct a bounded operator, $Q_{m}=\oint_{\mathcal{S}_{2}}F$ which will be $1$ on the unglued slice by the Bianchi identities. 

However, we can not use the above argument in the glued slice and the same operator $Q_{e}$ for the wormhole background will not be zero. In fact $\mathcal{S}_{d-1}$ will be a non contractible cycle in the theory and there will be non trivial flux coming out of each mouth of the wormhole. There will be a new operator that is present in the glued slice, which is the Wilson loop operator that goes across the handle and comes back to the starting point through $M$. If we call this loop $\gamma_{\star}$, since $\gamma_\star\cdot \mathcal{S}_{d-1}=1$,
$\big[\oint_{\gamma_\star}A,\,Q_e\big]=i$, so $Q_e\in\mathbb Z$, and for $d=3$,
$Q_m\in 2\pi\mathbb Z$ because of Dirac quantization. For the orientation of the cycles such that normal vectors are pointing away from the mouths, we have $Q_{ej/m,1}=-Q_{e/m,2}$. These will be central elements to our algebra operators and gives us different superselection sectors of the theory. In particular, we have for any region $U$ in the common part of the two
slices that contains such a shell,
$\mathcal A_{T=\infty}(U)\cong\mathcal A_{W,T}(U)/(Q_e=Q_m=0)$. This issue arises only in the case the neck is added onto a single universe. Additional modifications will also follow if we have higher form gauge fields. In particular for instance if one have a gauge field $B$ with a $d$ form field strength $H=dB$, we may consider an operator $e^{i\oint_{\mathcal{S}_{d-1}}B}$ localized in the same annular region. This however can be written as a bounded function of a smeared operator in the infinitely long tube $e^{i\int_{N_{i}}H}$\footnote{Compactifying the asymptotically cylindrical end by including its boundary is the same as including the points $p_{i}$ in the balls, where the gauge fields are smooth.}, which is equal to one in the presence of no matter. However, in the glued geometry the cycle $\mathcal{S}_{d-1}$ is not contractible and we can not write it in terms of a bounded function of smeared $H$ field. This is an new additional operator that we would have to include.   

As we have been discussing in the previous sections, we will construct the states in our theory using path integration to the past of our chosen Cauchy slice, with boundary conditions given by the above identification. However, here we perform the standard Lorentizan QFT rather than the gravitational path integral. In particular, we are considering two Cauchy slices, corresponding to the two fixed backgrounds. The first Cauchy slice has no wormholes but two finite tubes of length $T$. The boundary conditions for our path integral, will be the restriction of the field configurations on the Cauchy slice in our original theory we started with, up to the conformal blow up we have preformed in each of the two balls. In the region $M$ we have the same background metric for the spacelike slice and we have the same set of boundary configurations for the path integral.  

Similarly for the Cauchy slice with a handle, in the region $M$, we have the same background metric and boundary field configurations. However in the wormhole neck region, the field configurations follow the above conditions we specified, i.e, \eqref{con1},\eqref{con2}. We consider the state constructed in such way with no boundary operator insertions. This defines positive linear forms, $\omega_{T}$ and $\omega_{W,T}$ on each of the $*$ algebras discussed above. Following this we can construct the Hilbert space of states by the GNS construction. As usual, this is done by imagining $\mathcal{A}_{T}$ as a vector space and that for any two operators $\Phi_{1}$ and $\Phi_{2}$ in $\mathcal{A}_{T}$, we have the corresponding vectors states with an dinner product given by,
 \begin{equation}
     \langle\Phi_{1}|\Phi_{2}\rangle =\omega_{T}(\Phi_{1}^{\dagger}\Phi_{2}).
 \end{equation}
Thus, the states are defined up to an equivalence relation $\ket{\Phi+X}\sim \ket{\Phi}$ as long as $\omega_{T}(X^{\dagger}X)=0$. After taking the completion of the vector space of these states with the above inner product one gets the associated Hilbert space $\mathcal{H}_{T}$. Similar discussion as above follows for the state $\omega_{W,T}$ and algebra $\mathcal{A}_{W,T}$ which will lead to a distinct Hilbert space $\mathcal{H}_{W,T}$.

We have said that the states above, $\omega_{T=\infty}=\omega$ and $\omega_{W,T}$, are constructed using the path integral to the past of each of the slices, $\Sigma$ and $\Sigma_{T}$. We would like to compare these two states to see what the IMP gluing would look like for the background (seed) states. For that we need to understand in detail the properties of these states for regions far away from the balls where the surgery happens, and the handle. From theorem 1 of \cite{Isenberg:2001qe} for four spacetime dimensions and its generalization to higher dimensions, we see that the metric of the glued slices approaches arbitrary close to the original metric in the regions far away from the neck and balls as we take $T$ to be big. In general the estimate is 
\begin{equation}\label{metest}
    |g_{T}-g| \leq Ce^{-aT/2}\sim C\big(\frac{\rho_{h}}{R}\big)^{a}
\end{equation}
where $C$ is a constant and $a$ is a positive constant and is for example $1/2$ in four dimensions.

To see how this translates to the states $\omega$ and $\omega_{T}$, we note that states prepared by path integrals are expected to have the right short distance singularity behavior. In particular, in order to have a Hilbert space of `reasonable size', i.e, at most countably infinite dimensions, we need to put the states of the high energy modes (and the low energy modes for open universes) in their ground states. This behavior is summarized by statement, all well behaved states look like the vacuum at short distances. (For the low energy modes, the appropriate asymptotic behavior the spacetimes dictate how the modes behave. For asymptotically AdS for instance, the condition implies the low energy modes are in their ground state, i.e, the state they would have in the AdS vacuum.) Generally speaking this nice short distance behavior is satisfied by states prepared by the path integral, see \cite{Hartle:1976tp,Hack:2012qf}. Therefore, we assume here too the states $\omega$ and $\omega_{T}$ are also such states.

States with this sort of correct short distance property are called Hadamard states, and they generally have two basic structures, a global singular terms that captures the needed singularity behavior, called the Hadamard parametrix and a state dependent smooth part where the unique content of the state is present. That is two point function of local operators (usually called a bi-distribution) is given by,
\begin{equation}
    H(x,y)= \frac{\Gamma(\frac{d-1}{2})}{4\pi^{d+1/2}}\bigg[\frac{U(x,y)}{d_{\epsilon}(x,y)^{(d-1)/2}}+V(x,y) \text{ log }(d_{\epsilon}/l^{2})\bigg] + w(x,y)
\end{equation}
where $d_{\epsilon}(x,y)= \sigma(x,t)+2i\epsilon(t_{x}-t_{y})+\epsilon^{2}$, and $\sigma(x,y)$ is the squared geodesic distance between close enough points $x$ and $y$, while $l$ is some arbitrary length scale. The functions $U(x,y)$ and $w(x,y)$ are smooth functions of x and y with $U(x,x)=1$. Similarly, $V(x,y)$ is also a smooth function that is only non zero for even spacetime dimensions, $d+1$. The universal singular behavior is captured by the first two terms whereas $w$ is the state dependent smooth part of a Hadamard state.  One usually writes,
\begin{equation}\label{v}
    V(x,y) = \sum_{k=0}^{\infty} v_{k}(x,y)\sigma^{k}
\end{equation}
where $v_{k}$ are again smooth functions\footnote{This series is usually asymptotic however, one can achieve pointwise convergence using the appropriate cutoffs \cite{friedlander1975wave,Hollands:2001nf}}. 

Thus both the path integral states $\omega$ and $\omega_{T}$ will have two point functions like $H(x,y)$ above with their own coefficients $U,v_{k}$ and the geodesics distances computed in their respective metrics. For a free field theory, this two point function completely determines the state, since all its higher point functions are given in terms of the two point function, $H(x,y)$ by Wick contractions. The computation of these object is straight forward and it follows from requiring that the above two point function satisfies the equations of motion for fixed $y$. One then compares the coefficients of $\sigma^{k}$ order by order leading to a set differential equations \cite{garabedian1964partial,Riesz1949}. These equations depend on the background and can be solved recursively. If one wants to put a given QFT on different backgrounds, the dependence of these equation will only be in the background metric. In other words, for our case of a sequence of metrics $g_{T}$ that gets close to $g$ as \eqref{metest}, these coefficients also approach one another. 

One can also express $w$ like \eqref{v}, with its coefficients $w_{k}$. These coefficients however are not completely determined by the theory and the background metric, rather, as we have stated earlier, they are unique to the particular state chosen and correspond to the physical content of the state. Any well behaved state will have these same singular behaviors and differ in the smooth part $w$\footnote{One usually only needs to specify the $0^{th}$ order coefficient $w_{0}$, the rest of the higher order coefficients can be uniquely determined from the equations using $w_{0}$.}.   

Coming back to our states $\omega$ and $\omega_{T}$, since they are well behaved states, their singular part is that of Hadamard. In addition, as $g_{T}$ approaches $g$, the coefficients of the singular terms approach one another. However, for the smooth parts of the states, it is in general not true that they match. In particular, the presence of a handle (or a neck connecting two a prior disconnected universe) will introduce more correlation between points separated by long distances. Since the geometry is significantly modified deep in the bulk, it would be reasonable to expect that the smooth parts of the states constructed by a path integral (with no action of additional operators) would be different. However, for regions far away from the handle, where the metric is close to the original one, one may suspect that the states would produce the same correlators. To be slightly more precise, at very long distances, the wormhole `throat' area can not be resolved and the handle would look like a line connecting two points $p_{1}$ and $p_{2}$. The effect of the throat at these long distances is that it will act like an electric or magnetic source for a possible gauge field in our theory. Thus we say that the two states produce the same correlations when restricted to their respective algebra of operators $\mathcal{A}_{T=\infty}(U)$ and $\mathcal{A}_{W,T}(U)/(Q_{e}=Q_{m}=0)$, for any region $U \subset M$ far away from the wormhole neck. The fact that this is true is for the state prepared by path integral depends on the the fact that the wormhole background looks like the original background, at long distances, with magnetic and/or electric sources acting at points $p_{1,2}$. 

Alternatively, we do not have to use the path integral to construct the states, at least the state of the wormhole background. Rather, we can construct this state by `hand'; first considering the universal singular part to correspond to that of a Hadamard state for the background metric $g_{T}$. On the other hand we can take the smooth part to be constructed with some smooth cutoff functions by gluing the smooth part of the original state for regions far away from the handle, and the smooth part of any other state (for example the one constructed by path integral with the wormhole background being the future boundary) for regions near the handle. This procedure gives another Hadamard state since the equations that $w$ satisfies are given locally by the equations of motion, similar to $U$ and $v_{k}$. Therefore, by construction the states produce the same correlation functions for operators in regions far away from the neck. This can be taken as the quantum version of the IMP gluing. 

What we have said so far is about the free theory. For interacting field theories, the higher point functions are not determined by the two point functions of the Hadamard state. Even for the two point functions, the above determination of the coefficients is not sufficient as there would be addition terms in the equations of motion that depend on the interaction coupling. There would be addition expansion parameter, the coupling constant and one has to match order by order in both expansion parameters, $\sigma$ and $\lambda$ to recover the coefficients $v_{k}^{(n)}$, which now depend on $n$, the order in the $\lambda$ expansion. These again only differ by the terms involving the metric for a given theory defined on different backgrounds metrics \cite{Hollands:2001nf,Hollands:2001fb}. 

\subsection{ER=EPR, gravitational interactions and all that}\label{er=epr}

Let's come back to the question we asked before, what precise new element is added in the case where a wormhole handle is added. The above discussion implies that the algebra of operators and correlation functions in well behaved states with correcting high energy singularities far way from the wormhole handle can be made arbitrarily the same as the correlation functions in the original background, modulo the electric and magnetic charges induced by the wormhole mouths. This is the charge without charge first discussed in \cite{Wheeler:1957mu}. This is almost identical, in IR limit, to the situation where one removes two points from a simply connected slice, modifying both the first and the second homology groups, and imposing the appropriate boundary conditions. However, this similarity between the two situations stops when we look at the geodesic completeness of the two slices, the one with a wormhole handle and the one with two punctures, since only the slice with a wormhole handle is geodesically complete. This point is related to the main point we are trying to make, which is that the new element (in addition to the charges) that emerges by the wormhole gluing is a throat, $H$ (see figure \ref{fig2}) on the volume maximizing $\tau=0$ slice, an invariant way to divide the full (or part of the) slice.

We started out with conformally blowing up to ball regions into a asymptotically cylindrical regions. Then we divided the asymptotically cylindrical ends in to three finite subregions and another infinite subregion. The wormhole hand is constructed by removing the infinitely long subregions and identifying the points on the three finite subregions in each of the ends, after a certain twisting of the coordinates. The subregions close to the mouths of each of the tubes correspond to the regions $E_{1,2}$ of the wormhole handle, while the middle finite subregions correspond to the $E_{0}$ region of the wormhole. As we can see from the discussion in section \ref{usecom}, the horizon is located at $l=0$ in the $E_{0}$ subregion. 

We can consider the subregion of $E_{0}$ to the left and to the right of $l=0$, and construct the von Neumann algebra of operators associated with the subregions, after thickening. These algebras of operators are are generally expected to be type III$_{1}$ and have infinite amount of entanglement with each other and  with the algebra associated to the rest of the slice. This follows from the standard story in quantum field theory in curved spacetime, because of what we discussed above, every Hadamard state have the same high energy behavior and the entanglement between the two subregions is governed by the high energy modes very close to the horizon, which therefore looks like the Rindler wedge. It is known that Rindler wedges have infinite amount of entanglement across the Rindler horizons \cite{bisognano1975duality,bisognano1976duality,OAI:Scale-and-conformal-symmetry-in-four-dimensional-operational-quantum-field-theory-September-26-2026}. However, this argument seems to lead us to the conclusion that there is no fundamentally new element added here since, one has the same amount and type of entanglement between the two subregions of $E_{0}$ both in the wormhole slice and in the original slice with asymptotically cylindrical ends. It seems like locally there is no new feature added due to the gluing of a handle that does not modify regions far away, and the `connection' across the throat is the same as any other `connection' across an imaginary dividing line (in the present context, a sphere) in a smooth spacelike slice. The only new feature for the wormhole case, however, is the emergence of this dividing sphere can be uniquely specified and is an invariant notion, under metric deformations.

We recall that the metric in the $E_{0}$ region of the glued slice is 
\begin{align}
      g_{T}=2^{4/d-2}R^{2}e^{-T} \;\text{ cosh}^{4/d-2}(l(d-2)/2)\;\big(dl^{2}+d\Omega_{d-1}\big) + \cdots
\end{align}

In the presence of a negative cosmological constant, we can choose the slice where the gluing happens to be a time symmetric one, not just $\tau=0$, which we do here to simplify our computations. In this case, the ingoing and outgoing null expansions are given by
\begin{align}
    \theta_{\pm} &= \pm\frac{1}{A} \frac{dA}{ds}\; \text{ where, } \frac{ds}{dl}= 2^{2/d-2}Re^{-T/2} \;\text{ cosh}^{2/d-2}(l(d-2)/2):=a\\
    \text{ and, }& A=\frac{2\pi^{d/2}}{\Gamma(d/2)}\;a^{d-1}
\end{align}
We denoted the area of the co-dimension 2 spacelike sphere by $A\propto a^{d-1}$. Then we have
\begin{equation}
    \frac{da}{dl} = a \; \text{tanh}\big(\frac{l(d-2)}{2}\big), \text{ and, } \frac{dA}{ds} = \frac{2\pi^{d/2}}{\Gamma(d/2)}(d-1)a^{d-2}\text{ tanh}\big(\frac{l(d-2)}{2}\big)
\end{equation}
Thus we find that,
\begin{equation}
    \theta_{\pm}=\pm \frac{1}{A}\frac{dA}{ds}= \pm\frac{d-1}{a}\text{ tanh}\big(\frac{l(d-2)}{2}\big)
\end{equation}
As we would expect from our earlier computations in section \ref{usecom}, the marginally outer trapped surface, $H$, is at $l=0$, where $\theta_{-}$ also vanishes. 

Consider now a small metric perturbation that would modify the sphere at the horizon by $\xi$ in any direction, $v^{i}$, that is orthogonal to the sphere. This then modifies the null expansions of sphere by an elliptic operator acting on $\xi$, 
\begin{equation}
    \delta_{\xi} \theta = \theta[\xi]-\theta[0]=(-\Delta_{S} - R_{ij}v^{i}v^{j})\xi,
\end{equation}
as discussed very nicely by the authors of \cite{Andersson:2007fh,Andersson:2005gq}. $R_{ij}$ above is the Ricci tensor of the slice. The contraction $R_{ij}v^{i}v^{j}$ is similar to the contraction one would use when deriving the Hamiltonian constraint equation, where one writes the contraction of the Einstein tensor by the normal vector to the Cauchy slice and write it in terms of the extrinsic curvature and the Ricci scalar of the Cauchy slice itself which is then related to the matter content on the slice (see for example \cite{Wald:1984rg}). Using the same relation here but for the Cauchy slice and the sphere in question, we have
\begin{equation}
   2 R_{ij}v^{i}v^{j} = R_{\Sigma} -R_{S} + \pi_{ab}\pi^{ab} + \pi^{2}
\end{equation}
where $\pi_{ab}$ is the extrinsic curvature of the round sphere and $\pi$ is its trace. On the other hand $R_{S}$ is the Ricci scalar of the sphere. Thus we have in the $E_{0}$ region for large $T$,
\begin{equation}
    R_{ij}v^{i}v^{j}\approx\frac{1}{2}R_{\Sigma} -\frac{1}{2}R_{S} =\Lambda - \frac{(d-1)(d-2)}{2\rho_{h}^{2}} + 8\pi G_{N}\rho
\end{equation}
Thus $ -R_{ij}v^{i}v^{j}$ is positive as long as $G_{N}$ is perturbatively small or zero, so is the operator $-\Delta_{S}$ as we have mentioned previously for a different purpose, we can see that if we were to require $\theta[\xi]=0$ to find the horizon, we can just invert $-\Delta_{S} - R_{ij}v^{i}v^{j}$ to get a unique $\xi$. Thus we have a unique surface that is defined by $\theta[\xi]=0$ that is stable under small metric perturbations.

The point of the above discussion is to elaborate on the question, how should we understand the slogan ER=EPR? In particular on the ER$\implies$EPR direction of the equality. Since the same kind of entanglement is present across any other imaginary line for a well behaved state. Is the statement only about a two (possibly disconnected) regions are now connected by a smooth surface? The argument above implies we should rather say ER implies the existence of an invariant (under small metric perturbations) notion of a dividing surface (in this case a sphere), which has infinite amount of entanglement between its left and right hand sides, in the limit the Newton's constant $G_{N}=0$.     

How should we then proceed to including gravitational perturbations by turning on $G_{N}$. First note that for the above approximation to work $\xi$ has to be small. In particular, we have $\delta_{\xi}\theta << 1/\rho_{h}$ and if we have  $(-\Delta_{S} - R_{ij}v^{i}v^{j})^{-1} << L$, then we would have
$
    \xi<<\frac{L}{\rho_{h}}
$.
But we only have $(-\Delta_{S} - R_{ij}v^{i}v^{j})>(d-1)(d-2)/2\rho_{h}^{2}$, thus $(-\Delta_{S} - R_{ij}v^{i}v^{j})^{-1} <\frac{2\rho_{h}^{2}}{(d-1)(d-2)}$. This means we still have 
\begin{equation}
    \xi<<\frac{2\rho_{h}}{(d-1)(d-2)}.
\end{equation}
Thus as long $\rho_{h}>> l_{P} \sim G_{N}^{1/(d-1)}$, $H$ is stable under perturbative gravitational corrections. We say more on this and related points on section \ref{hibstat}.

The IMP gluing in the quantum theory does not change the correlators far away from the surgery area but introduces a stable (under quantum gravitational corrections) dividing surface $H$ where the subregions of $E_{0}$ to the left and and to the right of $H$ correspond to type III$_{1}$ von Neumann algebras of operators, and therefore have the entanglement pattern associated with those algebras, when $G_{N}=0$. On the other hand, for the other direction of ER=EPR, it would be more precise to say that rather than just `entanglement', entanglement (of the type III$_{1}$ von Neumann algebra's kind for $G_{N}=0$) across a surface that is stable under small metric perturbations  like $H$, equals the Einstein-Rosen bridge, since as we said before entanglement of this type is present across any co-dimension 2 spacelike surface of a smooth spacetime.  

\subsubsection{Perturbative corrections}

To proceed further with perturbative corrections, we first note the importance of the stable co-dimension 2 surface, $H$. One of the main issues with applying the operator algebraic tools used in algebraic QFT, to perturbative quantum gravity is that naturally in QFT one associates bounded functions of the (smeared and local) operators of our theory to any open subset of the full spacetime. These correspond to the observables that one can measure localized in those open subsets of the background spacetime. However, for perturbative quantum gravity these open subsets are are hard to unambiguously define since the spacetime itself fluctuates, unless one `anchors' to a region where gravitational interactions vanish and can be specified without ambiguity like the boundary of quantum gravitational theories defined on asymptotically flat or asymptotically AdS spacetime. Alternatively, one can define observables relationally with respect to the experimenter measuring the observables like in \cite{Aguilar-Gutierrez:2023odp, Witten:2023qsv,Chandrasekaran:2022cip,Witten:2023xze}  to mention few among many, or a feature of the state that is stable under the perturbative gravitational fluctuations of the spacetime like in \cite{Bahiru:2023zlc,Bahiru:2022oas}. Our situation is convenient for the later kind of description. 

We thus define the algebra of observables by first gravitationally dressing \cite{Donnelly:2015hta,Donnelly:2016rvo} smeared local operators with respect to $H$ (on the volume maximizing slice with $\tau=0$ which fully specifies it with the condition $\theta[H]=0$) by shooting a geodesic perpendicular to $H$, and requiring them to act at a maximum geodesic distance $a\sim O(1)$ from $H$, in either increasing $l$ (right side operators) or decreasing $l$ direction (left side operators). We construct the left side algebra $\mathcal{A}_{L}$ and right side algebra $\mathcal{A}_{R}$, by considering bounded functions of the left side and the right side operators. Note that this procedure in impossible to in the original spacetimes and are only possible due to the wormhole gluing we have done. That is this is a new feature due to the gluing. 

In the limit $G_{N}=0$, these operators reduce to the operators that we construct in the QFT on the $E_{0}$ region of the wormhole background. Thus, as we have stated above, after taking the weak closure, $\mathcal{A}_{L}$ and $\mathcal{A}_{R}$ are type III$_{1}$ von Neumann algebras. These are algebras of operators in the regions very close to the surface $H$, in the `middle' of the the wormhole handle, whose background metric is given by a metric that is approximately AdS-Schwarzschild in the large $T$ limit, see section \ref{usecom}. We can choose the state on $\Sigma$ to be a Hartle-Hawking like state close to the wormhole throat,  which is approximately symmetric under the generator of the Killing vector field, $\chi$, that preserves the two exterior regions of AdS-Schwarzschild metric and $H$. We therefore consider the following operator,
\begin{equation}
    h = \int_{\Sigma}d^{d}x\;  g(x) \; \chi^{\mu}n^{\nu}\;T_{\mu\nu}
\end{equation}
where $g(x)$ is a smooth smearing function strictly localized in the region $E_{0}$, a geodesic distance $a$ from $H$. The behavior of $\chi$ is not specified since its use is only to define $h$ and even though $h$ defined on all of $\Sigma$, it does not act on the rest of the background but $E_{0}$; and it preserves the $E_{0}$ region. Such an operator is not an element of neither $\mathcal{A}_{L}$ nor $\mathcal{A}_{R}$. But we may consider an operator like 
\begin{equation}
    h_{R} = \int_{H_{+}}d^{d}x\;  g(x) \; \chi^{\mu}n^{\nu}\;T_{\mu\nu}
\end{equation}
to be a candidate to be an element of of $\mathcal{A}_{R}$. Here by $H_{+}$ we mean the region to the right of $H$ and excluding the region to the left of $H$. It is clear that this operator acts like $h$ by generating translations along the vector field $\chi$ for operators in $\mathcal{A}_{R}$. This operator however has the usual $G_{N}$ dependent divergences due to the particular limit we are taking. In particular, we have derived in section \ref{usecom} that $\rho_{h}^{d-1} \sim G_{N}M \propto (Re^{-T/2})^{d-2}$. When we take the perturbative limit $G_{N}\rightarrow 0$, we implicitly assumed that we are keeping $Re^{-T/2}$ fixed, in other words we have $M$  scaling like $c/G_{N}$ in this limit, for a constant $c$. This contribution produces a divergence for the above operator $h_{R}$. For $h$, this divergence is absent because the Killing vector field $\chi$ generates the opposite translation in the left side from the right side. We therefore redefine $h_{R}$ as $h_{R}-M$ to remove the problematic divergence and we see that the new $h_{R}$ can indeed be affiliated to $\mathcal{A}_{R}$, that is its bounded functions can be included into the algebra of operators dressed to the right of $H$. This operator, acting only on the right side of $H$, causes a time shift between the two mouths, visible for any observer far in the region $M$. Therefore, it is a quantum gravitational operator mapping a given state on a given background to a state on a different background, i.e, one with a time shift between the mouths.

Therefore including perturbative correction in $G_{N}$ to $\mathcal{A}_{R}$ will requires us to gravitationally dress the operators with respect to $H$ and include the new $h_{R}$ to the algebra. For this kind of gravitational dressed operator $O$, we have, 
\begin{equation}
    [h_{R},O] = [h,O] = -i\partial_{t}O
\end{equation}
That is bounded functions of $h_{R}$, like $e^{ixh_{R}}$ generate an outer automorphisms of the algebra $\mathcal{A}_{R}$ acting like $e^{ix(h+X)}$, where now $X$ commutes with any of the operators in $\mathcal{A}_{R}$. As the operator $h$ acts exactly like the boost operator of the Rindler wedge (which is the modular Hamiltonian in this case) close to $H$, taking the closure of this set of observables $\{\mathcal{A}_{R}, e^{ix(h+X)}\}$ will give us a crossed product algebra \cite{Takesaki1973,Witten:2021unn},
\begin{equation}
    \mathcal{A}_{R}\rtimes \mathbb{R}_{X},
\end{equation}
which is a type II$_{\infty}$ von Neumann algebra.

Thus we conclude, from the above arguments, that ER=EPR, when $G_{N}$ is turned on perturbatively, is more precisely stated as, ER is the same as an entanglement pattern of type II$_{\infty}$ von Neumann algebras across a stable co-dimension 2 surface $H$. 

\subsubsection{Beyond perturbative corrections}

Can we continue this discussion to finite $G_{N}$? The first problem to do this would be the fact that now $H$ is not a stable surface. We have specified that $H$ is stable under gravitational perturbations as long as $\rho_{h} >> l_{P}$. We take $G_{N}$ finite, i.e, on the order of a finite radius of the horizon $\rho_{h}$, then $\xi$ produced by gravitational perturbations will be large enough to modify $H$. Alternatively, we can take $\rho_{h}$ to be on the order of $l_{P}$, while still keeping modes away from the wormhole region much bigger than the Planck scale. For convenience, we take this later perspective in the following.  

Since we have $\rho_{h}\sim Re^{-T/2}$, we need to be careful on the about the order of limits to recover a Planck scale wormhole. Before we first took $G_{N}$ to zero, keeping $T$ fixed then took large $T$. However now we need to take $T\sim \text{ log }\frac{R}{G_{N}^{1/(d-1)}}=\text{ log }\frac{R}{l_{P}}$, so that $\rho_{h}/l_P$ stays fixed.   

In either of the cases discussed above, we will not have a unique and stable surface like $H$, more precisely it is quantum mechanically ambiguous, that is it has $O(1)$ quantum uncertainty (see section \ref{hibstat}). Therefore the algebras of operators that we have defined above do not make sense. First since we do not have a stable surface, under fluctuations, dressing the operators with respect to a surface like $H$ is not possible. Secondly, the condition we have set to define $H$, i.e, $\theta[H]=0$ can not be put since, as we will see later, $\Delta (\rho_{h}\theta) = O(1)$, which translates to an uncertainty in the position $l=l_{S}$ of any candidate co-dimension 2 surface like $H$. This uncertainty is again $O(1)$ and does not depend on $T$. 

Before continuing our discussion let's only consider the case where the wormhole connects two apriori disconnected universes. Given the issues we have mentioned above, we need to modify the questions we ask to understand ER=EPR more carefully. In particular we can not define $\mathcal{A}_{R,L}$ as above. Rather we define the algebra semiclassical operator by considering dressing the operators with respect to the boundary, note that the modes in the region away from the wormhole handle are treated semiclassically. Due to the strong curvature of the handle region, we also have an uncertainty width on where a `candidate' $H$ can be, which we denote by $\Delta s=s_+ -s_-$. Thus we introduce the right algebra of operators, $\mathcal{A}_{s_+}$ which includes operators dressed to the boundary and $s>ks_+$ for $k>>1$, that is from a distance at least $ks_+$ away from the highly curved region all the way to the boundary. On the other hand the left algebra of operators, $\mathcal{A}_{s_-}$, which is defined in a similar way with $s<ks_-$. The effect of an ER bridge of this kind is to produce a highly curved and quantum mechanical buffer zone for these two algebras of operators.

To say more is slightly difficult without knowing how to describe the Planck size wormhole. However, a possible way forward would be to consider the algebra of operators $B(\mathcal{H}_{0})$, where $\mathcal{H}_{0}$ is the boundary CFT Hilbert space sector that describes this specific configuration. By this we mean the sector where we take the boundary central charge ($N^{2}$ for the $\mathcal{N}=4$ SYM in 4d with $SU(N)$ gauge group) to be very large but finite, and the operators correspond to the bulk fluctuations that act far away from the wormhole handle correspond to multi trace operators with a conformal dimension $\Delta_{far}<<N^{2}$ and do not depend on $N$. The operators that act deep in the bulk corresponding to fluctuations acting on the quantum mechanical wormhole correspond to multi trace operators with conformal dimension $\Delta_{near} \sim l_{AdS}/l_P$. Because we have a dual description in the boundary, we know that there is some state $\ket{\Omega}$ defined at the boundary of the Cauchy slice $\partial\Sigma$ that corresponds to the bulk state with the highly quantum wormhole and the ambient slice we described. We do not have an obvious candidate for the state $\ket{\Omega}$, but we prepare it using the bulk gravitational path integral with the corresponding boundary condition on the $\tau=0$ slice, in particular $T\sim \text{log }\frac{R}{l_P}$, see section \ref{states}. We also know that $\mathcal{A}_{s_{+},s_{-}} \in B(\mathcal{H}_{0})$ (more precisely this is about the boundary duals of $\mathcal{A}_{s_{+},s_{-}}$, which are algebras only approximately, since we are considering large but finite $N$). In addition, we can consider the commutant of $\mathcal{A}_{s_{-}}$ inside $ B(\mathcal{H}_{0})$, for instance. In this case, we have 
\begin{equation}
    \mathcal{A}_{s_{+}}\subset \mathcal{A}_{s_{-}}^{'}.
\end{equation}

A useful point here is that, this construction is different from the situation where the boundary theories are dual to disconnected but thermally entangled AdS universes with a gas of gravitons, at a temperature below the Hawking-{age temperature. This is because here we see from the construction that $\mathcal{A}_{s_{+}}^{'} \wedge \mathcal{A}_{s_{-}}^{'} \neq C\mathbb{I}$, for the largest possible algebras  $\mathcal{A}_{s_{\pm}}$. This is not the case is not the case for disconnected but thermally entangled universes. If $\mathcal{A}_{l,r}$ are the largest possible algebras of operators for the left and right universes, in the semiclassical limit, the full Hilbert space is factorized as $\mathcal{H}_{l}\otimes \mathcal{H}_{r}$, and the algebras of operators are 
\begin{equation}
    \mathcal{A}_{r,l}\cong B(\mathcal{H}_{r,l}),
\end{equation}

where $\mathcal{H}_{l,r}$ are the low energy large $N$ sectors of the full Hilbert space. In particular, we can construct a non trivial intermediate algebra $\mathcal{N}_{R}$ between $ \mathcal{A}_{s_{+}}$ and $\mathcal{A}_{s_{-}}^{'}$. More precisely, the disconnected universes case corresponds to what is called a singular inclusion of $\mathcal{A}_{r}\subset \mathcal{A}_{l}^{'}$, while the Planck sized wormhole is not. The extreme case for the quantum wormhole is $\mathcal{A}_{s_{+}}^{'} \wedge \mathcal{A}_{s_{-}}^{'} $ is infinite and that the state $\ket{\Omega}$ is cyclic vector for it (that is one can generate all of $\mathcal{H}$ by acting with operators in $\mathcal{A}_{s_{+}}^{'} \wedge \mathcal{A}_{s_{-}}^{'} $). An intermediate case is that $\mathcal{A}_{s_{+}}^{'} \wedge \mathcal{A}_{s_{-}}^{'} $ is non trivial but $\ket{\Omega}$ is not a cyclic vector for it.  In the extreme case, with the additional assumption of split property for $\mathcal{A}_{s_{+}}$ and $ \mathcal{A}_{s_{-}}$, we find that $\mathcal{N}_{R}$ is a type I$_{\infty}$ factor that can be canonically constructed with the modular conjugation operator, $J$, of $\mathcal{A}_{s_{+}}^{'} \wedge \mathcal{A}_{s_{-}}^{'} $ with respect to $\ket{\Omega}$, as
\begin{equation}
    \mathcal{N}_{R}= \mathcal{A}_{s_{+}} \vee J\mathcal{A}_{s_{+}}J
\end{equation}
Similarly for $\mathcal{N}_{L}$, which includes $\mathcal{A}_{s_{-}}$. 

If however, one were to extend the low energy sector of the left and the right CFT's by including heavy operators with $\Delta \sim l_{AdS}/l_P$ (that is, for instance, $N$ dependent but scale slower than $N^{2}$ for the $\mathcal{N}=4$ SYM example), our configuration would correspond to these heavy operators acting in the left (at $s_{-}$) and the right (at $s_+$) universes and being entangled. This then realizes ER=EPR beyond perturbation theory. We believe to add more concert comments concerning $\mathcal{A}_{s_{+}}^{'} \wedge \mathcal{A}_{s_{-}}^{'} $, we need a theory of the Planck sized wormhole which is currently not available.

\section{Gravitational perturbation theory}\label{section5}

\subsection{The perturbative Hilbert space}\label{hibstat}

To construct a Hilbert space for the set of states that we discussed in \ref{states}, there are a couple of properties required. These were discussed in \cite{Witten:2022xxp}. The first property is that one has to define a sesquilinear inner product between any two states in the set\footnote{To be a bit more precise the addition of inner product structure only gives us a pre-Hilbert space. One gets the full Hilbert space after a closure of the set of states in question with respect to the prescribed inner product.}. After defining the Hilbert space, one has to also remember that we preformed a BRST quantization and therefore, the physical space of states is the cohomology class the above set for the BRST charge operator given in subsection \ref{boundgaufix}. 

A point we have not emphasized in the previous sections is that the states are not only functions of the conformal class of Reimannian metrics on $\Sigma$, $\Sigma$ being the $\tau=0$ slice, but also the several ghost and antighost fields, $c^{\mu}$ and $\bar{c}^{\mu}$ on the slice. In addition, in the presence of matter fields, the wave function will also depend on the matter degrees of freedom on $\Sigma$. However, there is a term in the action that is supposed to fix $\Sigma$ to be the $\tau = 0$ slice. This involves the antighost field $\bar{c}^{0}$ and after the action of the BRST differential we see that it looks like,
\begin{equation}
    S_{gf,0} = \int_{\Sigma} d^{d}x\sqrt{\text{det }g}\bigg( \varphi \tau + \bar{c}^{0}\mathcal{I}c^{0} \bigg)
\end{equation}
where $\varphi$ is the auxiliary field associated with $\bar{c}^{0}$ and doing the path integration over $\varphi$ gives the condition $\tau = 0$. On the other hand, the variation of $\tau$ leads to 
\begin{equation}\label{i}
    \mathcal{I} =  -\Delta_{g} + K_{ij}K^{ij} + 8\pi G_{N} \hat{T}_{00}
\end{equation}
where $\hat{T}_{00} = T_{00} - \frac{g_{00}}{d-1}T^{\mu}_{\mu}$, $T_{\mu\nu}$ here is the matter stress-energy tensor including the contribution from cosmological constant (in \cite{Witten:2022xxp} $\mathcal{I}$ is denoted by $\Xi$)\footnote{In perturbation theory, it can be seen that $\mathcal{I}$ is a positive operator and therefore there will not be any issues that would arise from the presence of zero modes, note that the contribution from $\hat{T}_{00}$ is perturbatively small and thus will not affect the positivity of $-\Delta_{g}$.}. 

A few things can be said about $\mathcal{I}$. The first is $\Delta_{g}= g^{ij}D_{i}D_{j}$, and it acts on a wave function that depends on the conformal class of the metrics on $\Sigma$. The path integral is over the space of metric fluctuations that change the conformal class of $g_{0}$. If we take some $g_{c}$ to be a representative of the conformal class, one has 
\begin{equation}
    \Delta_{g} = \phi^{q}\;g_{c}^{ij}D_{i}D_{j}
\end{equation}
$\phi$ is now an operator which, following the Licnerowicz equation \eqref{lichno}, we can write in terms of the conformal class representative and the extrinsic curvature by a field re-definition. Coming to the extrinsic curvature, recall that $\tau = 0$ and thus $\Pi_{ij} =  \frac{\sqrt{\text{det }g}}{8\pi G_{N}}K_{ij}$, thus one has
\begin{equation}\label{K}
    K_{ij} = \sigma_{ij} +\frac{8\pi G_{N}}{\sqrt{\text{det }g}} \frac{\delta}{\delta g^{ij}}.
\end{equation}
The first term in the above equation is the classical value of the extrinsic curvature. Therefore, the effect coming from the $K_{ij}K^{ij}$ terms in \eqref{i}, when $O(G_{N})$ corrections are included, was interpreted by \cite{Witten:2022xxp} as a $T^{2}$ deformation of the classical limit of the above action $S_{gf,0}$ (albeit the auxiliary field is integrated out). 

Doing the path integral over $\bar{c}^{0}$ will bring down the determinant of $\mathcal{I}$ and impose the condition, $c^{0} = 0$, that was alluded to earlier.

The operator $\mathcal{I}$ will have the following expansion
\begin{equation}
    \mathcal{I} = -\Delta_{g}+\sigma_{ij}\sigma^{ij}+ 8\pi G_{N}\; \mathcal{I}_{1} + (8\pi G_{N})^{2}\; \mathcal{I}_{2}
\end{equation}
where $\mathcal{I}_{1}(.)= \frac{\sigma^{ij}}{\sqrt{\text{det }g}} \frac{\delta}{\delta g^{ij}} ( .) + \frac{1}{\sqrt{\text{det }g}} \frac{\delta}{\delta g^{ij}} (\sigma^{ij}\; .) +\hat{T}_{00}\; (.)$, and $\mathcal{I}_{2} = \frac{1}{\text{det }g} \frac{\delta}{(\delta g_{ij})}\frac{\delta}{(\delta g^{ij})}.$

Therefore we have, 
\begin{align}
    \text{det } \mathcal{I} &= \text{det }\bigg(-\Delta_{g}+\sigma_{ij}\sigma^{ij}+ 8\pi G_{N}\; \mathcal{I}_{1} + (8\pi G_{N})^{2}\; \mathcal{I}_{2}\bigg)\\
    &=  \text{det } (-\Delta_{g}+\sigma_{ij}\sigma^{ij}) \bigg(1+  8\pi G_{N}\;\text{Tr }((-\Delta_{g}+\sigma_{ij}\sigma^{ij})^{-1} \; \mathcal{I}_{1})  + O(G_{N}^{2})\bigg)
\end{align}

Then the inner product between two states is
\begin{align}\label{innpro}
    \langle \psi_{1}|\psi_{2}\rangle_{1} &= \int_{\text{fields on } \Sigma} D[g]Dc^{\mu}D\bar{c}^{\mu} \;\psi_{1}^{\star}([g],\tau=0,\cdots) \;e^{iS_{gf,0}}\;\psi_{2}([g],\tau=0,\cdots)\\ 
    &= \int_{\text{fields on } \Sigma} D[g]Dc^{i}D\bar{c}^{i} \;\psi_{1}^{\star}([g],0,\cdots) \; (\text{det }\mathcal{I}) \; \psi_{2}([g],0,\cdots)
\end{align}
The Hilbert space is defined from the set of the perturbative states constructed in the previous section with respect to this inner product. As we mentioned before, the physical Hilbert space is only the cohomology class of this Hilbert space we have just constructed. The BRST cohomology condition will remove the dependence of the wave functions on the ghost and antighost modes, in addition to imposing diffeomorphism invariance on $\Sigma$, i.e, the wave functions depend on the conformal class $[g]$ up to diffeomorphisms. We discuss this condition together with the algebra of operators acting on the physical Hilbert space in the following subsection.   

However before preceding to the algebra of operators, we want derive the claim that was made in the previous section that $\Delta(\rho_h \theta)=O(1)$, when the horizon radius, $\rho_h$ is close $l_P$. First we note that on the $\tau=0$ slice, set by our gauge fixing choice, the null expansion $\theta$ is given by
\begin{equation}\label{theta}
    \theta = \frac{1}{A}\frac{dA}{ds} - K_{ij}v^{i}v^{j}
\end{equation}
where $v^{i}$ are unit normal to the co-dimension two surface. We have assumed in our discussions that the extrinsic curvature of $\Sigma$ at the classical level is zero, for the sake of simplicity of our computations. However, at the quantum level (as we turn on $G_{N}$ perturbatively), we will not have that luxury. In particular, following the discussion around \eqref{K} we have,
\begin{equation}
    K_{ij}=\frac{8\pi G_{N}}{\sqrt{\text{det}g}}\frac{\delta}{\delta g^{ij}}
\end{equation}
even when $\sigma_{ij}=0$. Thus the extra term in \eqref{theta} is indeed needed. In the previous section, we have computed that in the $E_{0}$ region we have,
\begin{equation}
   x:= \frac{1}{A}\frac{dA}{ds} = \frac{d-1}{a}\text{tanh}\big(\frac{l(d-2)}{2}\big), \text{ and } a \propto Re^{-T/2}\text{cosh}^{2/d-2}\big(\frac{l(d-2)}{2}\big)
\end{equation}
In other words,
\begin{equation}
    \rho_hx = O(1)
\end{equation}
and does not depend on $G_{N}$, as $\rho_h\sim Re^{-T/2}$.

On the other hand, as usual the symplectic form on $\Sigma$ (check \cite{Witten:2022xxp} to see how the phase space changes in the context of the conformal method of treating the constraint equations),
\begin{equation}
    \Omega = \int_{\Sigma}\delta \Pi^{ij}\delta g_{ij}
\end{equation}
implies the commutation relation which leads to 
\begin{equation}
    (\Delta g_{ij})(\Delta \frac{\delta}{\delta g_{ij}}) \approx1.
\end{equation}
Close to the region of the horizon, we have $\Delta g_{ij} \sim \rho_h^{2}$ and $\sqrt{\text{det}g}\sim \rho_h^{d}$. Thus raising the indices of $K_{ij}$ and noting that the components $v^{i}\sim 1/\rho_h$ too, we see that 
\begin{equation}
    \Delta(\rho_h \theta) \sim \rho_h\frac{G_{N}}{\rho_h^{d}}\rho_h^{4}.\frac{1}{\rho_h^{4}} = \frac{G_{N}}{\rho_h^{d-1}}.
\end{equation}
We therefore conclude that as $\rho_h$ becomes very close to the Planck length $l_P$, $\Delta(\rho_h \theta)= O(1)$.

\subsection{The algebra of operators}

Here we discuss the algebra of operators on the full Cauchy slice as a whole, for both the slice with a wormhole handle and one without. In the earlier section, we discussed the wormhole region and the far away region more closely, focusing on the unique features due to the wormhole gluing. There are a couple of points that needs to be clarified in the above inner product above. The first point is that the inner product is not positive definite, particularly because it involves path integrals over the ghost and anitghosts. In addition, the states should depend on the conformal class of the metric modulo diffeomorphisms so the inner product should reflect that. Both of these points are closely related to the fact that the states that we should be considering are really the ones in the cohomology class of $Q$. To understand these issues and the perturbation theory better, we construct the algebra of operators that act on the physical Hilbert space, $\mathcal{H}_{\text{phy},\Sigma} = \frac{\text{ker }Q}{\text{im }Q}$. We apply the discussion that follows for the set of states associated with both $\Sigma$ and $\Sigma_{T}$ to construct the physical Hilbert space associated with each.

The set of states in the previous section, after taking the closure with respect to the inner product \eqref{innpro} form a Hilbert space, $\mathcal{H}$. The wave functions will be functions of the ghost and antighost fields, $c^{i},\bar{c}^{i}$, the matter fields (if they exist) and the conformal class of the metric $[g]$. Thus the algebra of operators that act on this space includes bounded functions of the matter and ghost/antighost fields. In addition, there will be metric fluctuations that modify the conformal class of the metric. From the construction, however, Weyl transformations or transformations that do not change the conformal class of the metric are identified with the identity. This in fact shows the non perturbative `flavor' of the construction in that it is not only perturbative fluctuations which do not change the conformal class that are identified with identity, but also large changes within the conformal class are equivalent to the identity operator. In other words, the wave functions are functions on the space of conformal structures on $\Sigma$ or $\Sigma_{T}$ and the operators acting on these functions are bounded functions on the space of conformal structures. Let's call this algebra of operators $\mathcal{A}$ (or $\mathcal{A}_{T}$ which would be acting on $\mathcal{H}_{T}$ associated with $\Sigma_{T}$).

The algebra $\mathcal{A}$ will be the full algebra of operators and the physical algebra of operators should be the sub algebra of BRST invariant operators. To this end, we consider $\delta_{Q}$ with the property that $\delta_{Q}^{2}=0,$ that acts on the algebra of operators, $\delta_{Q}(\mathcal{A}) \subset \mathcal{A}$ such that $$\delta_{F}(\delta_{Q}(a)) =  1 + \delta_{F}(a), \text{ mod }2.$$
with $\delta_{F}$ being the fermion number\footnote{This operator generally is called a graded differentiation, in this particular case with a $\mathbb{Z}_{2}$ grading.}. As usual, $\delta_{Q}$ acts with an (anti)commutation with $Q$ \eqref{Q}, i.e,
\begin{equation}
    \delta_{Q}(a) = Qa - (-1)^{\delta_{F}(a)}\;aQ
\end{equation}
If we let $\mathcal{A}_{\text{ker}}$ to be the kernel of $\delta_{Q}$ in $\mathcal{A}$, then it is easy to see that $\mathcal{A}_{\text{ker}}$ is a $*-$ algebra. If $a,b \in \mathcal{A}_{\text{ker}}$ then
\begin{equation}
    \delta_{Q}(ab)= \delta_{Q}(a)\;b+ (-1)^{\delta_{F}(a)}a\;\delta_{Q}(b) = 0
\end{equation}
so $ab \in \mathcal{A}_{\text{ker}}$; and the involution acts like\footnote{We see that $$ (\delta_{Q}(a)^{*})^{*} = -(-1)^{\delta_{F}(a)}(\delta_{Q}(a^{*}))^{*} =  (-1)^{2\delta_{F}(a)} \delta_{Q}(a) = \delta_{Q}(a).$$}
\begin{equation}
    \delta_{Q}(a)^{*} = -(-1)^{\delta_{F}(a)}\delta_{Q}(a^{*}).
\end{equation}
Since $\delta_{Q}^{2}=0$, we have $\delta_{Q}(\mathcal{A}) \subset \mathcal{A}_{\text{ker}}$ and $\delta_{Q}(\mathcal{A})$ is an (two sided) ideal of $\mathcal{A}_{\text{ker}}$. Thus BRST invariant operators would be given by 
\begin{equation}
    \mathcal{A}_{0}= \frac{\mathcal{A}_{\text{ker}}}{\delta_{Q}(\mathcal{A})}
\end{equation}
Looking at the $G_{N}$ expansion of the inner product in the previous section and since our concern is with a perturbation theory, we can similarly describe the algebra of operators in perturbative expansion. A general operator in our algebra will have an expansion in the formal power series in $G_{N}$, of operators that do not contain $G_{N}$ explicitly. That is,
\begin{equation}
    a = \sum_{n} (G_{N})^{n}a_{n}
\end{equation}
where all the $a_{n}$'s have the same fermionic number. In addition, one can think of the BRST charge as 
\begin{equation}
    Q=Q_{0}+G_{N}Q_{1}
\end{equation}
for linear operators $Q_{0}$ and $Q_{1}$ acting on $\mathcal{H}$. Therefore we will also have,
\begin{equation}
    \delta_{Q} = \delta_{Q,0} + G_{N}\delta_{Q,1}.
\end{equation}
We expect for the physical algebra of operators to have a faithful representation on the physical Hilbert space. The BRST charge $Q$, with $Q^{2}=0$, implements $\delta_{Q}$ on $\mathcal{H}$, and the physical space of states is the cohomology class of $Q$. To make this space into a Hilbert space however, one needs to define a `compatible' inner product that is independent of the representative for the equivalence class of the states. For instance the inner product \eqref{innpro} is in general different for different choices of representatives in the cohomology class of the states $\psi_{1,2}$. To define such a compatible inner product we would like Q to be Hermitian with respect to this same inner product, that is we need to modify \eqref{innpro} so that,
\begin{equation}
    \langle\psi|Q\phi\rangle_{\eta} = \langle Q\psi|\phi\rangle_{\eta},
\end{equation}
If we define a Hermitian involution operator $\eta$ is such that $Q^{\dagger} = \eta Q \eta^{-1}$\footnote{In Krein space theory an operator such as $\eta$ is known as fundamental symmetry operator.} where $Q^{\dagger}$ is the adjoint with respect to the original inner product. We then have,
\begin{equation}
    \langle\psi|\phi\rangle_{\eta} = \langle\psi|\eta\phi\rangle_{1} = \int_{\text{fields on } \Sigma} D[g]Dc^{i}D\bar{c}^{i} \;\psi^{\star} \; (\text{det }\mathcal{I}) \; \eta\phi
\end{equation}
For convenience we will not carry the subscript $\eta$ of the inner product in the discussions that follow. We now consider a sub space $\mathcal{H}_{\text{ker}}$ of $\mathcal{H}$, which is $Q^{-1}(0)$, and its sub space $Q\mathcal{H} \subset \mathcal{H}_{\text{ker}}$.

Thus we define the physical Hilbert space as $\frac{\mathcal{H}_{\text{ker}}}{Q\mathcal{H}}$ with the inner product on the equivalence classes as 
\begin{equation}\label{equvinn}
    \langle [\psi]|[\phi]\rangle := \langle\psi + Q\alpha|\phi + Q\beta\rangle_{\eta}
\end{equation}
where $\psi,\phi \in \mathcal{H}_{\text{ker}}$, while $ [\psi],[\phi]$ are equivalence classes associated with each states respectively; in addition $\alpha,\beta$ any states in $\mathcal{H}$ therefore $\psi + Q\alpha$ and $\phi + Q\beta$ are any representatives from $ [\psi],[\phi]$ equivalence classes respectively.  

The above inner product is well defined, i.e, it is independent of the choice of representative from the equivalence class, precisely because of the Hermiticity and nilpotency of $Q$, since the terms involving $Q$ in the expansion of the left hand side of \eqref{equvinn} are all zero, i.e,
\begin{equation}
    \langle Q\alpha|\phi\rangle_{\eta} = \langle \psi|Q\beta\rangle_{\eta} =  \langle Q\alpha|Q\beta\rangle_{\eta} = 0.
\end{equation}

The operator $\eta$ essentially plays the role of removing the contributions of the nonphysical degrees of freedom (ghost, antighosts and pure gauge modes) from the inner product so that the inner product does not depend on the choice of representatives from the equivalence classes. 

The physical algebra of operators act on the physical Hilbert space as 
\begin{equation}
    \pi(a)\ket{[\psi]} = \ket{[a\psi]}. 
\end{equation}
This representation \cite{Duetsch:1998hf} is well defined on the Hilbert space of equivalence classes if its action is BRST invariant, i.e, $$(a+\delta_{Q}b)\ket{\psi+Q\alpha} - a\ket{\psi}$$ is a null vector for $a \in \text{ker }\delta_{Q} $ and $ \ket{\psi} \in \mathcal{H}_{\text{ker}}$, while $\alpha $ is any state in  $\mathcal{H}$ and $b$ is any operator in $\mathcal{A}$. But note that since $\delta_{Q}a = 0$, we have $aQ = (-1)^{\delta_{F}(a)}Qa$; thus the terms that do not vanish because of the condition $Q^{2}=0$, are all of the form $Q\ket{\beta}$ for some $\ket{\beta} \in \mathcal{H}$. 

Taking the weak closure of the above representation $\pi(\mathcal{A}_{\text{phy},0})$ with respect to the physical inner product, one gets the von Neumann algebra 
\begin{equation}
    \mathcal{A}_{\text{phy}} =\pi(\mathcal{A}_{\text{phy},0})^{''}.
    \end{equation}

\appendix

\section{Elliptic operators}\label{ellpequ}

We discuss the definition and some facts about elliptic operators, an interested reader can find much more technical details in appendix II of \cite{Choquet-Bruhat:2009xil}, or \cite{Witten:2018lgb}.

If $(\Sigma, g)$ is a smooth Riemannian $d$-manifold, a linear differential operator $L$ of order $m$ is a map between sections of a tensor bundle on $\Sigma$ to sections of another tensor bundle,
\begin{equation}
    L : \mathcal{T}_{i_1\cdots i_S}(\Sigma) \rightarrow \mathcal{T}_{k_1\cdots k_N}(\Sigma) 
\end{equation}
of the form
\begin{equation}\label{linopp}
    Lu:=\sum_{r=0}^{m}a_{r} D^{r}u.
\end{equation}
where $ D^{r}$ denotes the covariant derivatives of order $r$. The tensor components in local coordinate are $(Lu)_{k_1\cdots k_{N}} \in  \mathcal{T}_{k_1 \cdots k_{N}}(\Sigma) $, for a smooth $u_{i_1\cdots i_{S}} \in \mathcal{T}_{i_1\cdots i_{S}}(\Sigma)$. The coefficients $\left(a_r\right)^{j_1\cdots j_r\, i_1\cdots i_S}{}_{k_1\cdots k_N}$ are also smooth functions on $\Sigma$. 

However for general non-linear operators, the coefficients $a_{r}$ depend not only on the point $x \in \Sigma$ but also on $u$ and its covariant derivatives, $D^{l}u$. An operator is said to be semilinear if $a_{m}$, the coefficient of the highest order derivative, depends only on $x$ while the rest of the coefficients, $a_{r<m}$, depend on $D^{l}u$ for $l<r$. On the other hand, an operator is called quasilinear if $a_{m}$ also depends on $D^{l}u$ for $l< m$ while it is still linear in $D^{m}u$. Otherwise, the operator is called fully nonlinear. 

The term in the equation \eqref{linopp} with the highest number of covariant derivatives is called the principal part of $L$,
\begin{equation}
   \left(a_m\right)^{j_1\cdots j_m\, i_1\cdots i_S}{}_{k_1\cdots k_N}D_{j_{1}}\ldots D_{j_{m}}
\end{equation}
The principal/leading symbol, $\sigma_{x}(.)$, of $L$ at point $x$ is given by the contraction of $a_{m}(x)$ with $\otimes^{m} (ip)$ where $p \in T^{*}|_{x}(\Sigma)$ are the momenta; in other words, it is given by replacing each covariant derivative of the principal part of $L$ with the imaginary unit times the momentum, $p$. Then the leading symbol at each point in $\Sigma$ is a map from the cotangent space at $x$ (homogeneous of degree $m$ in $p$) to linear maps between tangent spaces at $x$, that is,
\begin{equation}
    \sigma_{x}(p): T_{i_{1}\ldots i_{S}}|_{x}(\Sigma) \rightarrow  T_{i_{1}\ldots i_{N}}|_{x}(\Sigma)
\end{equation}
and in local coordinates,
\begin{equation}
    \sigma_{x}(p)^{i_1\cdots i_S}{}_{k_1\cdots k_N} = (i)^{m} \left(a_m(x)\right)^{j_1\cdots j_m\, i_1\cdots i_S}{}_{k_1\cdots k_N}\;p_{j_{1}}\ldots p_{j_{m}}
\end{equation}
We say the operator $L$ is underdetermined at the point $x$ if the map $\sigma_{x}(p)$, for all $p\neq0$, is surjective, while we call $L$ overdetermined at $x$ if $\sigma_{x}(p)$ is injective. We call $L$ elliptic at $x$ if $\sigma_{x}(p)$ is bijective or invertible, for all $p\neq0$. The most familiar examples are the Laplacian and the Dirac operators, which are both elliptic.

The leading symbol of the Laplacian, $\Delta=g^{ij}D_{i}D_{j}$, is given by
\begin{equation}
    \sigma_{x}(p)= - g^{ij}p_{i}p_{j} \leq0,
\end{equation}
and the leading symbol for the Dirac operator, $i\slashed{D} = i\gamma^{i}D_{i}$ is 
\begin{equation}
    \sigma_{x}(p)=\slashed{p}
\end{equation}
where $\gamma^{i}$'s are the gamma matrices. We see that both are non zero unless $p=0$, thus both the Laplacian and the Dirac operator are elliptic. A slightly more complicated operator that we use in the last appendix is the conformal Laplace operator,
\begin{equation}
   D_{i} (D^{i}Y^{j} + D^{j}Y^{i} - \frac{2}{d}D_{k}Y^{k}g^{ij}),
\end{equation}
for some vector field $Y^{i}$. It's leading symbol is given by
\begin{align}
    \sigma_{x}(p)&=-p_{i}p^{i} Y^{j}-p_{i}p^{j}Y^{i}+\frac{2}{d} p_{i}p_{k}Y^{k}g^{ij}\\
    &=-p_{j}p^{j}Y^{i}- \frac{d-2}{d}p_{j}p^{i}Y^{j}.
\end{align}
Thus $\sigma_{x}(p)=0$, for $p\neq 0$ and $Y^{i}\neq0$, implies that 
\begin{equation}
    p_{j}p^{j}Y^{i}Y_{i} + \frac{d-2}{d}p_{j}Y^{j}p^{i}Y_{i} =0
\end{equation}
which is impossible for $d\geq 2$. Thus the conformal Laplace operator is an elliptic operator for $d\geq 2$. In the next appendix, we review boundary conditions we need to impose in order to get a well defined perturbation theory, once we modify the relevant operators to be elliptic.

\section{Why conformal dynamical coordinates?}\label{b}

It is well known that gauge invariance creates certain problems for perturbation theory. Before gauge fixing a gauge invariant action will not give a well defined propagator or one -loop determinant and it is only after this issue is resolved that we are able to define perturbation theory. The simplest example would be the case of a massless $U(1)$ gauge field theory. Let $F = dA$\footnote{We assume we are in the topologically trivial sector of the gauge theory.} be the field strength and the action is given by
\begin{equation}
    S= \frac{-1}{4}\int_{M} d^{D}x\sqrt{\text{det }\hat{g}}\; F_{\mu\nu}F^{\mu\nu}
\end{equation}
To begin to define perturbation theory, we consider expanding $A$ around some classical solution as $A_{0}+a$, where $a$ is the perturbation, then to second order in $a$ the semiclassical theory will be described by the action 
\begin{equation}
    S = \frac{-1}{2}\int_{M} d^{D}x\sqrt{\text{det }\hat{g}}\; \bigg(\partial_{\mu}a_{\nu}\partial^{\mu}a^{\nu} - (\partial_{\mu}a^{\mu})^{2}\bigg) = \int_{M} d^{D}x\sqrt{\text{det }\hat{g}}\; a^{\mu}E_{\mu\nu}a^{\nu}
\end{equation}
This action does not have a well defined propagator since the operator $E_{\mu\nu}$ is non invertible and its leading symbol\footnote{Details concerning this and related ideas can be found in appendix \ref{ellpequ}} is $\sigma_{\mu\nu}(p) = p_{\mu}p_{\nu} - p^{2}\delta_{\mu\nu}$ has an infinite dimensional kernel, i.e, any vector proportional to $p_{\mu}$. Another way to see this issue would be to note that the gauge redundancy implies a gauge field $a$ is equivalent to $a_{\mu}+\partial_{\mu}\lambda$ for some function $\lambda(x)$. Therefore $\partial_{\mu}\lambda$ would be a zero mode for $E_{\mu\nu}$ since it would be equivalent to $a_{\mu}=0$. Of course the solution is to fix the gauge redundancy by picking a `good' gauge condition that would be equivalent to choosing a unique $\lambda$. Such a gauge condition for example would be $\partial_{\mu}a^{\mu}=0$ as can be seen from the form of the linearization $E_{\mu\nu}$.

Similar issues arise for gravity as we do the path integral above. We summarize the rather long linearized action for gravity \cite{Bastianelli:2013tsa} as (we do not include matter fields to simplify the discussion)
\begin{equation}
    S= \frac{-1}{8\pi G}\int_{\mathcal{M}_{+}} d^{D}x\sqrt{\text{det }\hat{g}}\;  \bigg(\hat{h}^{\mu\nu} (L\hat{h})_{\mu\nu} + \frac{1}{2}(D^{\nu}\hat{h}_{\mu\nu} - \frac{1}{2}\partial_{\mu}\hat{\mathrm{h}})^{2}\bigg) =  \frac{-1}{8\pi G}\int_{\mathcal{M}_{+}} d^{D}x\sqrt{\text{det }\hat{g}}\;  \hat{h}^{\mu\nu} (L^{'}\hat{h})_{\mu\nu}
\end{equation}
where $L$ involves the covariant derivative and curvature tensor of $\bar{g}$ and $\hat{\mathrm{h}}= \bar{g}^{\mu\nu}\hat{h}_{\mu\nu}$. The linearization $L^{'}$ has infinite dimensional zero modes, again because of the gauge redundancy and thus can not be inverted. However, as we have written it suggestively, one may get well defined perturbation theory if one sets the gauge condition,
\begin{equation}
    T_{\mu}(\hat{h})= D^{\nu}\hat{h}_{\mu\nu} - \frac{1}{2}\partial_{\mu}\hat{\mathrm{h}} =0 
\end{equation}
This gauge condition goes by many names some of which include de Donder or harmonic gauge. What we are essentially doing is modifying a non-elliptic operator, $L^{'}$, to an elliptic operator by fixing the gauge, which makes sure that the resulting operator has at most a finite dimensional kernel (and cokernel). To check whether this is a good gauge condition or not, we consider a diffeomorphism transformation $\hat{h} \rightarrow \hat{h}' = \hat{h} + \delta_{\epsilon}\hat{h} $ and check that $$T_{\mu}(\hat{h}+\delta_{\epsilon} \hat{h}) = 0$$ is satisfied for a unique $\epsilon_{\mu}$. This condition leads to an equation like $$(X\epsilon)_{\mu} = T_{\mu}(\hat{h}) $$ since $T_{\mu}$ is linear in $\hat{h}$, the above condition is equivalent to the question whether $X$ is invertible or not\footnote{More precisely, one has to check whether $T_{\mu}(\hat{h})$ is orthogonal to the zero modes of $X$ so that one can invert $X$ in the subspace orthogonal to its zero modes and solve for $\epsilon_{\mu}$.}.  

One can check that $L$ is elliptic (see appendix \ref{ellpequ}) and that its leading symbol is invertible, thus one can define a propagator in perturbation theory. One can also compute the one loop determinant however at this point an important problem arises, which has to do with $L$ being not positive definite\cite{PhysRev.160.1113,Gibbons:1978ac,Polchinski:1988ua}. The trace part of the metric fluctuation $\hat{h}$ has the wrong sign for its kinetic term and the one loop determinant will contain phase factor $i^{\infty}$ which is hard to interpret. There have been several discussions of this issue and it has also gain recent interest in relation to what an observer in a closed manifold might play to resolve the problem \cite{Maldacena:2024spf,Cotler:2025gui}. 

Until now we have been discussing some standard facts about perturbation theory and the problems posed by the existence of infinitely many zero modes, in which case the propagator and one loop determinant can not be found. This would be enough for an analysis concerning only closed manifold. However, for manifolds with boundary, as in the case that interests us, $\mathcal{M}_{+}$, there may be infinitely many zero modes arising from the boundary condition one imposes. For the case of gravity, from geometric analysis perspective, this is discussed in \cite{Anderson_2008}, and a physicist friendly discussion can be found in \cite{Witten:2018lgb}. 

Having removed the possibility of getting infinitely many zero modes in the bulk, one has to also make sure similar issues do not happen in regions very close of the boundary of the manifold. This is a local condition that has to be checked at each point on the boundary, thus we may as well approximate our manifold by $\mathbb{R}^{d}\cross \mathbb{R}_{+}$, and in this approximation we consider possible plane wave solutions for $$(L\hat{h})_{\mu\nu} =0$$ which in general are given by $e^{i\vec{p}.\vec{x} \pm i\omega x_{0}}$ where $x_{0}$ is the orthogonal direction to the boundary. For a solution of the above equation, locally, $\omega$ can be expressed in terms of the $\vec{p}$ using the leading symbol for $L$ and depending on the form of $L$\footnote{More precisely the form of the highest order terms in $L$, since those are the ones relevant for the leading symbol of $L$.}, $\omega$ can be real or complex. If it is complex (as it is for gravity), then one can get plane wave solutions along the boundary but an exponentially decaying or growing modes away from the boundary. Such modes will cause a problem for us particularly if they are decaying modes when $x_{0}$ increases, for sufficiently large $|\vec{p}|$, i.e, modes that decay like $e^{i\vec{p}.\vec{x} - \text{Im }\omega\; x_{0}}$. The reason is, in such cases, assuming $\text{Im }\omega$ grows with $|\vec{p}|$\footnote{Typically $\text{ Im }\omega(|\vec{p}|) \sim |\vec{p}|$ for elliptic operators.}, one gets almost zero modes localized very close to the boundary, $O(\frac{1}{\text{Im }\omega(\vec{p})})$ away from the boundary, for each arbitrarily large $|\vec{p}|$. Because of this accumulation of almost zero modes, $L$ can not have discrete spectrum and will have an ill-defined propagator for boundary conditions that allow for such decaying modes. Essentially our procedure will be to check that given a boundary condition, whether or not modes that decay exponentially as $x_{0}$ increases for arbitrarily large $|\vec{p}|$ exist. 

The boundary conditions associated with preparing the `naive' states, $\psi(g)$, will allow for such decaying modes and thus perturbation theory is ill-defined. To show this the first step is to note that the consistency of BRST quantization with the bulk gauge fixing $T_{\mu}(\hat{h}) =0$ (which will be discussed in the coming subsection) requires us to impose $T_{\mu}(h)|_{\Sigma} =0$ too.

We want to check if decaying plane wave solutions like $$\hat{h}_{\mu\nu} = \alpha_{\mu\nu}e^{i\vec{p}.\vec{x} - |\vec{p}|x_{0}}$$ are allowed by the boundary condition. As previously mentioned, the boundary condition for the naive states in the semiclassical limit can be stated as $\hat{h}_{ij}|_{\Sigma} = h_{ij} = 0$, which sets $\alpha_{ij} = 0$. However, one is left with boundary conditions for $h_{i0}$ and $h_{00}$ which will be dictated by the residual gauge fixing $T_{\mu}(h)|_{\Sigma} =0$. This will lead to four equations for $\alpha_{i0}$ and $\alpha_{00}$. It is easy to see that they look like,
\begin{align}
    ip^{i}\alpha_{i0}& - |\vec{p}|\frac{\alpha_{00}}{2} = 0\\
    |\vec{p}| \alpha^{i}_{0}& + i p^{i}\frac{\alpha_{00}}{2} = 0
\end{align}
and $\alpha^{i}_{0} = -\frac{ip^{i}}{2|\vec{p}|}\alpha_{00}$ is a solution. Therefore the problematic decaying solutions exist. In fact for any good gauge fixing condition one can impose, such modes always exist \cite{Anderson_2008}.

However, for the conformal boundary conditions we have the following; only the traceless part of $h_{ij}$ is going to be zero, therefore it remains to impose conditions for $\mathrm{h} = h^{i}_{i}$, $h_{i0}$ and $h_{00}$. There will be four equations coming from $T_{\mu}(h)|_{\Sigma}=0$ and one more from the addition condition on the boundary mean curvature.$$\hat{\tau}|_{\Sigma} = \tau =0.$$

We assume $x_{0}$ to be a coordinate parameterizing the geodesic distance from $\Sigma$ in the direction of $n^{\mu}$, which is the normal unit vector to $\Sigma$. Therefore, close to $\Sigma$ the metric looks like
\begin{equation}
    ds^{2} = -dx_{0}^{2} + g_{ij} dx^{i}dx^{j}
\end{equation}
In this coordinate system, the above condition on the mean curvature looks like
\begin{equation}
    \;\partial_{0}\mathrm{h} - 2 \partial^{i}h_{0i} + (\text{lower order terms in the derivative}) = 0
\end{equation}
Therefore, locally (or in the high momentum limit), we get the following totally five equations,
\begin{align}
    2ip^{i}\alpha_{i0}& + d|\vec{p}|\alpha = 0\\
     ip^{i}\alpha_{i0}& - |\vec{p}|\frac{\alpha_{00}}{2}+\frac{d}{2}|\vec{p}|\alpha=0\\
     |\vec{p}| \alpha^{i}_{0}& + i p^{i}\frac{\alpha_{00}}{2}+\frac{i(d-2)}{2}p^{i}\alpha=0
    \end{align}
where $\mathrm{h} = \delta^{i}_{i}\alpha \; e^{i\vec{p}.\vec{x} - |\vec{p}|x_{0}}$, and $\delta_{i}^{j}$ is the Kronecker delta symbol. These equations are all satisfied for non zero real $\vec{p}$ only if $\alpha_{i0} = \alpha_{00}=\alpha =0 $. Thus the problematic modes do not exist and this boundary condition leads to a well behaved perturbation theory. In addition to the availability of wormhole conformal surgery method, this is our main justification for using conformal dynamical coordinates.

\section{The conformal method}\label{confmeth}

We are particularly interested in the asymptotically AdS$_{D}$ spacetime, $\mathcal{M}$, and if one is given with a covariant theory with a well-posed initial value problem coupled with Einstein's gravity, then one can formulate a solution on the full spacetime from a set of initial data satisfying a certain constraint equations. The initial data will include a choice of a $d$-dimensional spacelike manifold $\Sigma \subset \mathcal{M}$ with a Riemannian metric $g$ and a symmetric 2-tensor $K$ (which would correspond to the the induced metric and second fundamental form on $\Sigma $, respectively), in addition there will be the rest of the non gravitational fields, collectively called $\Phi$, that are specified on $\Sigma$. The constraint equations are then given by \eqref{fullconequ}. In addition to the above gravitational constraints, there are also non gravitational constraints that we generally denote as $C(\Phi, g) = 0$. The first of these equations is called the momentum constraint and it imposes diffeomorphism invariance on the space of metrics on $\Sigma$ in the presence of matter, by producing the associated gravitational field to the matter content. On the other hand the second equation is called the Hamiltonian constraint. Roughly speaking, this will impose the condition that some metric on different spacelike $d$ manifold $\Sigma_{0}$ related to the original metric and slice $\Sigma$ by a translation with a timelike vector that goes to zero at spatial infinity, corresponds to the same initial data set. As such the region we will associate to the initial data set is the shaded region shown in figure \ref{fig1}, rather than just a single Cauchy slice with fields on top of it. To simplify some technicalities, we will assume this subregion is compact (by including the asymptotic infinity). One of the reasons for focusing on the asymptotically AdS$_{D}$ spacetimes is in fact this asymptotic boundary. Because the holographic and a non perturbative definition of quantum gravity is (or at least expected to be) a boundary quantum field theory for such spacetimes and the questions we will be asking, which may not have clear meaning generally in quantum gravity can be phrased in the precise language of the boundary theory. We will return to this point later when we discuss the quantum constraint equations.

As we mentioned before one important property of the classical constraint equations is that they are under-determined rather than elliptic, which means the equations do not determine the full data, rather there is a freely chosen part of the data and there is a part the will be determined by solving the constraints. But before talking about the above constraints in their full generality, we will make a few simplifications. The first of these simplification will be to consider no matter fields. The initial data sets in this case are called the vacuum initial data sets and they are the above constraint equations with $J_{k}=\rho = 0$. In addition, we will assume $\tau=\text{tr}_{g}K$ to be constant. We will keep this assumption in the subsequent sections too, since the gauge fixing that we will use in the perturbation theory will require the slices to be maximal, whose $\text{tr}_{g}K$ is indeed constant. Such initial data where $\tau$ is constant are called CMC (constant mean curvature) data. The reason for this separation of the extrinsic curvature into the trace and trace free part is first the fact the momentum constraint seems to simplify in the case that $\tau$ is constant, and that it is convenient in the application of what is called the conformal method (mainly developed by Licnerowicz, Choquet-Bruhat and York \cite{York:1972sj,PhysRevLett.26.1656,Friedrich1988,Friedrich:1998xmu,Beig:1993gt,Beig:1991dh,Friedrich:2003fq,Friedrich:2012iz}) in separating the free and determined part of the data. More technically the main point is to modify the constraint equations to an elliptic system of equations so that the solutions are exactly the determined initial data, and the conformal method implies that for a certain conformal fields the constraint equations are elliptic. Consider for example two metrics $g$ and $\tilde{g}$ related by a conformal transformation,
\begin{equation}
    \tilde{g}_{ij} = \phi^{q} g_{ij}
\end{equation}
for, $\phi>0$. Then it is not difficult to show that for any traceless symmetric tensor $\sigma$, we have $\tilde{D_{i}}\sigma^{ij} = 0$ if and only if $D_{i}(\phi^{\frac{(d+2)q}{2}}\sigma^{ij}) = 0$. On the other hand, for a convenient choice of $q= -\frac{4}{d-2}$ for $d>2$, the following relationship between Ricci scalars of the metrics holds;
\begin{equation}\label{tworicci}
    R(g) = \phi^{q-1} \Bigl ( R(\tilde{g}) \phi + (d-1)q \; \Delta_{\tilde{g}}\phi \Bigl )
\end{equation}
where $\Delta_{\tilde{g}} = \tilde{g}^{ij}\tilde{D}_{i}\tilde{D}_{j}$. Then for any symmetric traceless tensor that satisfies $\tilde{D_{i}}\sigma_{ij} = 0$, we can write an extrinsic curvature that satisfies the momentum constraint equations for $g$ (in the CMC-vacuum initial data case) as,
\begin{equation}\label{extricurv}
    K_{ij} = \phi^{-2} \sigma_{ij} + \frac{\tau}{d} g_{ij}
\end{equation}
Note that at this point there is no restriction on what $\phi$ can be, for any choice of $\phi$ and any choice of a symmetric and trace free 2-tensor whose divergence with respect to some arbitrary metric $\tilde{g}$ is zero, \eqref{extricurv} satisfies the momentum constraint with respect to $g$.

Following \eqref{tworicci} we can also say something about the Hamiltonian constraint. In particular, if we rewrite the terms involving $K$ in terms of $\sigma$ and $\tau$, 
\begin{equation}
  |K|^{2}_{g} - (\text{tr}_{g} K)^{2} = \phi^{qd} |\sigma|^{2}_{\tilde{g}} - \frac{d-1}{d}\tau^{2},  
\end{equation}
then the Hamiltonian constraint for $g_{ij}$ and $K_{ij}$ becomes the following constraint for $\phi$ and $\tilde{g}$,
\begin{equation}\label{lichno}
    \Delta_{\tilde{g}}\phi +\alpha \;\Bigl(R(\tilde{g})\phi - \; |\sigma|^{2}_{\tilde{g}}\phi^{\frac{2-3d}{d-2}}\Bigl) -\beta \phi^{\frac{d+2}{d-2}} =0, 
\end{equation}
where $\alpha = - \frac{d-2}{4(d-1)}$ and $\beta = \Bigl( \frac{d-2}{4d} \tau^{2} + \alpha\Lambda \Bigl)$. What the above equation is telling us is that now, the Hamiltonian constraint restricts what $\phi$ can be. In particular, $\phi$ should solve the above equation, called the Licnerowicz equation, so that the Hamiltonian constraint for $K_{ij}$ and $g_{ij}$ is satisfied. 

In summary, given some spacelike manifold $\Sigma$, if one picks an arbitrary metric $\tilde{g}$ and a symmetric and trace-less tensor, $\sigma$ which is divergence free with respect to $\tilde{g}$ on $\Sigma$, then \eqref{solconst} give initial data set that 
satisfy both the gravitational constraint equations (vacuum with constant mean curvature), if $\phi$ satisfies the Licnerowicz equation. The advantage of the Licnerowicz equation as opposed to the Hamiltonian constraint is that it is an elliptic equation and therefore $\phi$, the conformal factor, is exactly the part of the initial data that is fully determined by the constraint. Whereas the conformal class of $\tilde{g}$, a constant $\tau$ and a symmetric, traceless and divergence free tensor $\sigma$ are freely chosen. 

Finding $\sigma$, a divergence free two tensor with respect to $\tilde{g}$ on $\Sigma$, is relatively easy. The standard way is by using what is called York splitting, where we start with any symmetric traceless tensor $B_{ij}$ on $\Sigma$ and defining \begin{equation}\label{CY}
    C(Y)^{ij} = \tilde{D}^{i}Y^{j} + \tilde{D}^{j}Y^{i} - \frac{2}{d}\tilde{D}_{k}Y^{k}\tilde{g}^{ij},
\end{equation} for some vector field $Y^{i}$; then the tensor $$\sigma_{ij}:= B_{ij} + C(Y)_{ij}$$ is divergence free with respect to $\tilde{g}$ if and only if $\tilde{D}_{i}C(Y)^{ij}=-\tilde{D}_{i}B^{ij}$; and for any $B_{ij}$ a vector field $Y^{i}$ that solves that latter equation always exists, though not unique which does not change the definition of $\sigma_{ij}$.

When matter fields are present there will be a slight modification to the above form of the solutions to the constraints. There will be data, in addition to $\phi$, that will be determined by another elliptic equation, which is a vector field $W^{i}$. The $\sigma_{ij}$ present in \eqref{solconst} is then replaced by $\sigma_{ij} + C(W)_{ij}$ (where $C(W)^{ij}$ is the tensor defined on \eqref{CY} but now corresponding to the vector field $W^{i}$). The equation that $W^{i}$ satisfies will imply the momentum constraint is satisfied in the presence of matter fields and the Hamiltonian constraint is satisfied when the Licnerowicz equation is satisfied as before, but now with $\sigma_{ij}$ again replaced by $\sigma_{ij} + C(W)_{ij}$ and a term corresponding to the density of the matter fields is added.\footnote{One needs to be careful not to include interactions for the matter fields that would mess the nice properties of the Licnerowicz equation. More details in appendix \cite{chrusciel2018cauchy}}.
\begin{align}\label{solconstmod}
    g_{ij}&=\phi^{-q}\tilde{g}_{ij}\\
    K_{ij}&= \phi^{-2} (\sigma_{ij} + C(W)_{ij}) + \frac{\tau}{d}\phi^{-q}\tilde{g}_{ij}
\end{align}

Up till now the freely chosen data includes the conformal class of the metric, a constant mean curvature and as symmetric, traceless and divergence free tensor. One can also include the matter field (if we consider a single matter field with constraints for simplicity) into this data. These constraint equations generally depend on matter fields and the metric tensor on $\Sigma$. Therefore if we let $\tilde{\Phi}$ be part of the conformal data and we want $\Phi = \phi^{p}\tilde{\Phi}$ for some $p$ to be the physical matter field that satisfy the matter constraints with respect to $g$ (for simplicity we are considering here a single matter field with constraints), then if the constraints satisfy  
\begin{equation}\label{mattcons}
    C(\phi^{p}\tilde{\Phi}, \phi^{-q}\tilde{g}) = \phi^{r}C(\tilde{\Phi},\tilde{g})
\end{equation}
for some $r$, it would be enough to just require $C(\tilde{\Phi},\tilde{g}) = 0$ on the conformal data to make sure the physical constraints are satisfied. For the standard physically relevant theories like Einstein -Maxwell or Yang-Mills and others such a Weyl transformation exists \cite{Isenberg:2005xp}. More generally we want the action $F$ of the group of conformal factors on multiple matter fields to be such that $C(F(\tilde{\Phi},\phi), \phi^{-q}\tilde{g}) = \phi^{r}C(\tilde{\Phi},\tilde{g}) $, where $F(\tilde{\Phi},\phi) = \Phi$, $F(\tilde{\Phi},1) = \tilde{\Phi}$ and $F(F(\tilde{\Phi},\phi_{1}),\phi_{2}) = F(\tilde{\Phi},\phi_{1}\phi_{2})$. 

One usually solves for the physical matter fields as described here first, then move on to solving the momentum and Licnerowicz equations with the physical matter fields. Further discussion of the constraint equations in the presence of matter can be found in appendix \cite{chrusciel2018cauchy}, also check the last section of \cite{Witten:2022xxp} to see the treatment of higher form gauge fields and the presence of extra compact dimensions like the ones found in AdS/CFT. 

When the mean curvature is not constant, there will be slight modifications to the equations for $W^{i}$ and $\phi$ involving derivatives of $\tau$, however we are mostly concerned with the constant mean curvature scenario here. 

We have now reached at a general picture of the conformal method of solving for the classical constraint equations. It includes an arbitrary choice of metric tensor, symmetric TT (transverse(divergence free) and traceless) tensor, a constant mean curvature, which we take to be zero, and matter fields, ($\tilde{g},\sigma,\tau =0 ,\tilde{\Phi}$), on some spacelike manifold $\Sigma \subset \mathcal{M}$. If $\tilde{\Phi}$'s satisfy $C(\tilde{\Phi},\tilde{g}) = 0$, then $\Phi = \phi^{p}\tilde{\Phi}$ satisfies the matter constraints. Then, for some symmetric traceless tensor $\sigma + C(W)$, where the vector field $W$ is found by solving an elliptic equation, there is some $g$ and $K$ (when $\tau=tr_{g}K \neq 0$, we should say $K-tr_{g}K$ rather than just $K$) in the Weyl orbit of $\tilde{g}$ and $\sigma + C(W)$ respectively, such that $g$ and $K$ solve the Einstein constraint equations in the presence of matters fields $\Phi$. Therefore the conformal class of $g$ is freely chosen, but the conformal class of the $K$ is not freely chosen, rather the correct $K$ is in the conformal class of a special kind symmetric traceless tensors, $\sigma + C(W)$, where their longitudinal part is constructed out of $W^{i}$'s that solve the appropriate equation.

When matter is present, there will be a slight modification to the \eqref{solconst} since now the right hand side of the momentum constraint above does not vanish. Therefore now the solutions to the constraint equations, for the same set of freely chosen data as before, will be   
\begin{align}\label{solconst2}
    g_{ij}&=\phi^{-q}\tilde{g}_{ij}\\
    K_{ij}&= \phi^{-2} (\sigma_{ij} + C(W)_{ij} )+ \frac{\tau}{d}\phi^{-q}g_{ij}
\end{align}
where 
\begin{equation}
    \tilde{D}_{i}C(W)^{ij} = J^{j}
\end{equation}
and $C(W)^{ij} = \tilde{D}^{i}W^{j} + \tilde{D}^{j}W^{i} - \frac{2}{d}\tilde{D}_{k}W^{k}\tilde{g}^{ij}$ for some vector field $W^{i}$, and thus the above equation is again elliptic and $W^{i}$ is part of the determined data set.

\acknowledgments

I want to thank R. Mazzeo and D. Pollack for correspondences and Ki-Hong Lee for discussions. I want to thank Claude for quite useful comments on the draft, making the reference search efficient and the help on the latex code to draw the figures. This work is supported by the Israel Science Foundation (ISF) Grants No. 3191/23 and No. 2916/24, and by the United States-Israel Binational Science Foundation (BSF) Grant No. 2034142.




\bibliographystyle{JHEP}
\bibliography{biblio}
\end{document}